\documentstyle[aps,prb,preprint]{revtex}
\begin{document}
\draft 
\title{Non-linear response of a
 Kondo system: 
Perturbation approach to the 
time dependent Anderson impurity model}
\author{Y. Goldin and Y. Avishai\cite{AvishaiEM}}
\address{Physics Department, Ben Gurion University of the Negev\\
Beer Sheva, Israel}
\date{\today}
\maketitle
\begin{abstract}
Nonlinear tunneling current through a quantum dot (an Anderson 
impurity system) subject to both constant and alternating electric fields 
is studied in the Kondo regime. A systematic diagram 
technique is developed for perturbation study of the current 
in physical systems out of equilibrium governed by 
 time - dependent Hamiltonians of the 
Anderson and the Kondo models. The ensuing calculations 
prove to be too complicated for the Anderson model, 
and hence, a mapping on an effective Kondo problem 
is called for. This is achieved by 
constructing a time - dependent version of the 
Schrieffer - Wolff transformation. Perturbation 
expansion of the current is then 
carried out up to third order in the Kondo 
coupling $ J $ yielding a set of remarkably simple analytical 
expressions for the current. The zero - bias anomaly of the 
direct current ({\em DC}) differential conductance is shown to be 
suppressed by the alternating field while side peaks develop 
at finite source - drain voltage. Both the direct component 
and the first harmonics of the time - dependent response are equally 
enhanced due to the Kondo effect, while amplitudes of higher 
harmonics are shown to be relatively small. 
A ``zero alternating bias anomaly'' is 
found in the alternating current ({\em AC}) differential 
conductance, that is, it peaks 
around zero alternating bias. This peak is suppressed by the constant 
bias. No side peaks show up in the differential 
alternating - conductance but their 
counterpart is found in the derivative of the {\em AC}
with respect to the direct bias. The results pertaining to 
nonlinear response are shown to be valid also below the 
Kondo temperature. 
\end{abstract}
\newcommand{\beq}{\begin{equation}}
\newcommand{\eneq}{\end{equation}}
\newcommand{\bea}{\begin{eqnarray}}
\newcommand{\enea}{\end{eqnarray}}
\newcommand{\ak}{a_{k, \sigma}}
\newcommand{\akn}{a_{k}}
\newcommand{\akd}{a_{k, \sigma}^{\dagger}}
\newcommand{\aknd}{a_{k}^{\dagger}}
\newcommand{\bd}{b^{\dagger}}
\newcommand{\bp}{b_{p}}
\newcommand{\bpd}{b_{p}^{\dagger}}
\newcommand{\cms}{c_{m, \sigma}}
\newcommand{\cmsd}{c_{m, \sigma}^{\dagger}}
\newcommand{\cn}{c_{n}}
\newcommand{\cnd}{c_{n}^{\dagger}}
\newcommand{\cs}{c_{d, \sigma}}
\newcommand{\csd}{c_{d, \sigma}^{\dagger}}
\newcommand{\ea}{\epsilon_{1}}
\newcommand{\eo}{\epsilon_{1}}
\newcommand{\ed}{\epsilon_{d, \si}}
\newcommand{\ek}{\epsilon_{k}}
\newcommand{\eka}{\epsilon_{\ka}}
\newcommand{\ekp}{\epsilon_{k'}}
\newcommand{\ems}{\epsilon_{m, \si}}
\newcommand{\en}{\epsilon_{n}}
\newcommand{\ep}{\epsilon_{p}}
\newcommand{\eps}{\epsilon}
\newcommand{\epbr}{(\epsilon)}
\newcommand{\et}{\epsilon_{2}}
\newcommand{\fs}{f_{\sigma}}
\newcommand{\fsd}{f_{\sigma}^{\dagger}}
\newcommand{\gam}{\gamma}
\newcommand{\ia}{i_{1}}
\newcommand{\ib}{i_{2}}
\newcommand{\ic}{i_{3}}
\newcommand{\id}{i_{4}}
\newcommand{\ie}{i_{5}}
\newcommand{\ja}{j_{1}}
\newcommand{\jb}{j_{2}}
\newcommand{\jc}{j_{3}}
\newcommand{\jd}{j_{4}}
\newcommand{\je}{j_{5}}
\newcommand{\ka}{k_{1}}
\newcommand{\kb}{k_{2}}
\newcommand{\lam}{\lambda}
\newcommand{\nres}{\mbox{$ n_{\mbox{\scriptsize res}} $}}
\newcommand{\snres}{\mbox{\footnotesize $n_{\mbox{\tiny res}}$ 
}}
\newcommand{\nua}{\nu_{1}}
\newcommand{\nub}{\nu_{2}}
\newcommand{\ma}{m_{1}}
\newcommand{\mb}{m_{2}}
\newcommand{\mc}{m_{3}}
\newcommand{\md}{m_{4}}
\newcommand{\me}{m_{5}}
\newcommand{\om}{\omega}
\newcommand{\oma}{\omega_{1}}
\newcommand{\omb}{\omega_{2}}
\newcommand{\ombr}{(\omega)}
\newcommand{\qa}{q_{1}}
\newcommand{\qb}{q_{2}}
\newcommand{\si}{\sigma}
\newcommand{\sia}{\sigma_{1}}
\newcommand{\sib}{\sigma_{2}}
\newcommand{\ssa}{s_{1}}
\newcommand{\ssb}{s_{2}}
\newcommand{\ta}{t_{1}}
\newcommand{\tb}{t_{2}}
\newcommand{\tc}{t_{3}}
\newcommand{\tone}{t_{1}}
\newcommand{\ttwo}{t_{2}}
\newcommand{\tthree}{t_{3}}
\newcommand{\Ak}{A_{k, \si}(t)}
\newcommand{\DEpar}{\Delta E_{\parallel}}
\newcommand{\DEperp}{\Delta E_{\perp}}
\newcommand{\Gij}[1]{G^{#1}_{i, j}(t_{1},t)}
\newcommand{\Jkpk}{J_{k'k}(t)}
\newcommand{\Jt}{\tilde{J}_{k'k}}
\newcommand{\SM}{S(+\infty,-\infty)}
\newcommand{\SMr}{S(-\infty,+\infty)}
\newcommand{\SMp}{S_{p}}
\newcommand{\Tp}{\hat{T}_{p}}
\newcommand{\Vk}{V_{k d}(t)}
\newcommand{\Vkt}{\tilde{V}_{k d}}
\newcommand{\Vkta}{\tilde{V}_{\ka d}}
\newcommand{\Vktb}{\tilde{V}_{\kb d}}
\newcommand{\Wt}{\tilde{W}_{k'k}}
\newcommand{\Wkpk}{W_{k'k}(t)}

\section{Introduction}

An attractive research 
direction in contemporary condensed matter 
physics seems to be the study of non-equilibrium many - body 
phenomena. A promising experimental and theoretical 
framework for investigating this topic is 
provided by the physics of
quantum dot systems. The reason for that is clear, 
namely, quantum dots 
are fabricated and their properties can be 
elucidated by present day experimental techniques. 
Indeed, recent experiments on electron transport
in quantum dots at low temperatures
reveal signatures of coherent many - body physics 
such as the emergence of zero bias anomaly 
in current - voltage characteristics 
\cite{GoldhaberGordon98,Cronenwett98,Simmel98} 
which is due to the formation of a many body resonance. 
At the same time, the underlying theoretical models are
of sufficient simplicity, so that one encounters a 
rare occasion where one has an
experimentally accessible non-equilibrium quantum system
that is also amenable to reliable and controllable
theoretical approaches.

So far, the main effort in the physics of quantum dots
has been devoted to the study of  different phenomena emerging from 
the presence of large Coulomb interaction in the dot. 
In this contexts, the 
most familiar and simplest topic 
is the Coulomb blockade. Its 
essence is simply encoded as a
capacitance effect: every extra electron coming 
into the dot has to overcome a 
charging energy $ e^{2}/C $, where $ C $ is the 
capacitance of the dot. If the gate voltage 
is not tuned to supply this energy, 
the tunneling is (Coulomb) blocked. 
Yet, we believe that essentially genuine many - body 
aspects of resonant tunneling might be better revealed in the 
physics that goes substantially beyond the simplified Coulomb blockade 
picture. By this we mean the Kondo effect and other facets of strongly 
correlated electronic systems. They result from an intricate combination 
of electron - electron interaction and tunneling.

Whereas the Kondo effect in {\em bulk} materials 
has been a thought - inspiring subject 
of research for more than three decades (for review, see Ref. 
\onlinecite{Hewson:book93,Fulde:book91}), 
its emergence in {\em quantum dot physics} appears to be 
relatively new. 
Yet, it proves to be equally thought - inspiring. 
In particular, it opens a road to explore the 
{\em non-equilibrium} Kondo physics. 
Its hallmark is the zero - bias anomaly, that is, 
an appearance of a large 
narrow peak in the differential conductance around zero bias. 
Pertinent experiments have been carried out
on crossed - wire tungsten junctions \cite{Gregory92}, quenched 
lithographic point contacts \cite{RalphBuhrman92,RalphBuhrman94}, 
metal and metallic glass break junctions \cite{Keijsers9596,Yanson95} 
and, recently, quantum dots 
\cite{GoldhaberGordon98,Cronenwett98,Simmel98}. 

In parallel, progress has also 
been recorded in numerous theoretical works, 
\cite{Hershfield91,Kawabata91,MWLKondo93,Ng93,Yeyati93,%
MeirWingreenKondo94,HettlerSchoeller95,SivanWingreen96,%
KonigSchon96,Ng96,SchillerHershfield96,%
LangrethNordlander948,GoldinAvishai98,Lopez98}. Among other 
directions of research, interest is focused on time - dependent aspects of 
the Kondo physics 
\cite{HettlerSchoeller95,Ng96,SchillerHershfield96,%
LangrethNordlander948,GoldinAvishai98,Lopez98}. Although an 
actual experimental research has not yet been carried out, measurement 
of alternating tunneling current in the appropriate range of frequencies 
proves to be feasible 
\cite{Sollner83,Sollner84,Rydberg89,Brown8991}, 
\cite{comDipLayer}. Moreover, application of an external alternating 
electric field non-trivially affects the 
{\em dc}, an observable which 
can actually be measured 
\cite{KouwenhovenPRB94,KouwenhovenLett94,%
Blick95,Fujisawa97,Oosterkamp99}. We hope that the pertinent 
experiments can be carried out in the near future. 

It is useful to briefly mention the main features of 
the Kondo effect in bulk systems and in quantum dots. 
The Kondo effect was first revealed in the 
early 30's as an enhancement in the
resistivity of certain metals with decreasing temperature. Being a puzzle 
for thirty years it was then attributed by Kondo \cite{Kondo64} to 
interaction between conducting electrons 
and magnetic impurity atoms. Subsequent investigations
proved that at low temperatures a hybrid state is formed 
consisting of the conducting electrons assembled around an
impurity so that at zero temperature the 
magnetic moment of the impurity is completely screened. 

Within the realm of quantum dot physics, 
the impurity spin is represented by the spin 
of a single electron which is virtually locked in a 
deep level of the quantum dot. 
The formation of the hybrid state 
(either in bulk systems or in quantum dots) 
is accompanied by an appearance of a narrow peak in the 
interacting density of states of the 
impurity (or dot electron) close to the 
chemical potential of the band (lead) electrons. 
This peak is termed as an ``Abrikosov - Suhl 
resonance'' or ``Kondo resonance'' (see Ref. 
\onlinecite{Hewson:book93}, pp.109, 127 -- 132, 210). 
For quantum dots 
in a {\em static} non-equilibrium situation, when two chemical 
potentials are present in the system (one for every lead), 
the pertinent peak in the interacting density of states splits, 
under certain conditions, 
into two peaks, one at each chemical potential 
\cite{MWLKondo93,MeirWingreenKondo94,%
SivanWingreen96,KonigSchon96}, \cite{comTwoPeaks} 
(see Fig. \ref{Newfig1}a). 

The main object of our study in the present work is 
the response of a quantum dot to the application of a {\em strong time - 
dependent external electric field}. For simplicity it is assumed 
that the field is monochromatic, whose frequency $ \Omega $ is 
in the range of tens of gigahertz. 
The relevant Physics implies an interesting extension 
of the underlying Kondo physics since,
in a time - dependent field,
every eigenstate of electrons in the 
leads is split \cite{TienGordon63,Tucker79} into a 
family of states whose (quasi) energies are
separated by the photon energy $ \hbar \Omega $. 
Consequently, there is a family of Fermi seas, and, correspondingly, a 
family of chemical potentials in each lead. Thus, the Kondo peak in the 
density of states might split into a whole set of peaks 
(see Fig. \ref{Newfig1}b). Numerical calculations of 
the {\em DC} \cite{HettlerSchoeller95} indicate that this is 
indeed the case. We think that the occurrence of many separated 
Abrikosov-Suhl resonances can be tested experimentally. 
It will show up as side peaks in the (otherwise monotonic) 
$I(V)$ curve displaying the zero bias anomaly of the {\em direct} 
current. Another interesting 
question is how the 
contributions of these peaks to the current, which are mutually 
coherent, interfere. This interference 
can show up only in the {\em time - dependent} 
current \cite{comDipLayer}. 

In this work, both direct and alternating
tunneling currents are calculated within a perturbation 
scheme specially adapted for systems out of equilibrium. 
So far, calculations of the current through a Kondo system subject to a 
time - dependent bias were carried out using various assumptions and 
approximations. Here we carry out straightforward 
non-equilibrium perturbation expansion of the current in powers of the 
coupling constant between the quantum dot (or impurity atom) and the 
electrons in the conduction bands. Indeed, perturbation theory proved 
its usefulness in calculations related to
the equilibrium Kondo model in bulk systems.  
\cite{Kondo64,Appelbaum66,Appelbaum67}. 
Recently it has been used
 for quantum dots described by the Anderson model 
(in the Kondo regime) with a 
constant voltage bias \cite{SivanWingreen96}.

Let us then briefly list the main achievements of the present study.
Our formulation starts from the time - dependent Anderson model 
which has already been shown to adequately 
describe the essential physics of a quantum dot (or a tunneling system 
based on an impurity atom) at voltages and frequencies less than the 
level spacing in the tunneling region. We then develop 
the pertinent diagram perturbation technique which is able to 
treat non-equilibrium tunneling problems 
such as the time - dependent Anderson and Kondo models 
(for the former one it employs the slave - bosons method). 
A key point in the derivation is 
provided by combination of a specific approach suggested by Coleman 
\cite{Coleman84} to perform self consistent quantum and 
thermodynamical averaging  in strongly correlated systems, with 
the  Schwinger - Keldysh non-equilibrium Green functions 
formalism. As it turn out, calculations pertaining to the Anderson 
model appear to be 
rather cumbersome. At this stage we are therefore content 
by stating the rules for calculating diagrams and 
by pointing out a specific example 
where the Kondo behavior shows up. 
At the same time, we find it more practical to 
map the original time - dependent Anderson Hamiltonian
on an effective time - dependent Kondo
Hamiltonian. This is achieved by introducing a time 
dependent Schrieffer - Wolff transformation. 
Perturbation expansion of the current is then carried out within 
the Kondo model up to third order in the Kondo coupling $ J $ (sixth 
order in the tunneling coupling between conducting electrons and the 
dot). Remarkably simple analytical expressions are obtained for the 
whole spectrum of the tunneling current. 
It is shown that the zero - bias anomaly  of the {\em DC} 
differential conductance
is, in general, suppressed by an external alternating field, while
side peaks develop at higher source drain bias. 
The nonlinear time - dependent current is found to be an interference 
sum of ``direct - like'' contributions, 
each one with an effective bias determined
by the number of absorbed or emitted photons. The interference is 
shown to be rather destructive for all higher
harmonics except the direct and the first ones. 
These two harmonics (and only them) 
are enhanced as a result of the Kondo effect 
while the other harmonics remain relatively small. 
In this sense the Kondo system behaves like a usual resistor (although 
the current is Kondo enhanced). Namely, 
direct and first harmonic input voltage 
lead to a direct and first harmonic current response. 
This result is 
demonstrated to be remarkably different from that for a non - interacting 
one - level system where all the harmonics emerge together. 
Moreover, it is shown 
to be valid also below the Kondo temperature. 
Finally, we also calculate the differential alternating 
conductance and find that like its direct analog 
it displays a zero alternating - bias anomaly. 
Inspecting the dependence of the {\em AC}
 on the direct bias reveals 
a non - trivial structure marked by side peaks. 

The rest of the paper is organized as follows. In the next section, 
diagrammatic rules for the non-equilibrium perturbation expansion of 
the time - dependent Anderson model in the slave - boson 
representation are formulated. The diagrams are drawn 
up to sixth order in the tunneling coupling between the leads and the dot 
($ \Vkt $) and some formal analytical expressions for the current 
are obtained. Further elaboration of the perturbation expansion in this 
model meets computational problems which today look severe. In 
section \ref{TDSW} {\em a time - dependent version} of the Schrieffer 
- Wolff transformation is developed 
in order to transform the time dependent 
Anderson model onto a Kondo - type 
model. In section \ref{KondoPerturbation} perturbation expansion of 
the current (in the time - dependent Kondo model) is carried out 
in powers of the coupling strength ($ \Jt $). An 
analytical expression for the current is obtained and the novel results 
arising from that expression are discussed. The paper is concluded with 
a summary which, in particular, includes some prospects for further 
research directions. In view of the inherent complexity of the 
pertinent formulation, we try to present it in a pedagogical style.

\section{Perturbation expansion of the current 
in the time-dependent Anderson model}
\label{Anderson}
\subsection{Bare Hamiltonian and parameter specification}
In this section we first introduce the time - dependent Anderson 
Hamiltonian and specify the range of parameters 
appropriate for the pertinent physical problem. Then we  
recall the slave - boson approach 
to the Anderson model and combine it with the Schwinger - 
Keldysh non-equilibrium Green function formalism 
in order to get an equation for the tunneling 
current. Then we develop a perturbation expansion of the 
current in powers of the tunneling strength ($ \Vkt $) up to sixth order 
in which the Kondo physics above the Kondo temperature (or in 
a strong enough external field) is unraveled.

The time - dependent Anderson Hamiltonian takes the form, 
\beq                                      \label{AndersonModel}
   \tilde{H}_{A} = \sum_{k \in L,R; \si} \left( \ek + 
                                   \Delta_{L(R)}(t) \right) \akd \ak 
          + \sum_{\si} \ed \csd \cs  
          + \frac{1}{2} U \sum_{\si, \sigma' \ne \si} n_{\si} n_{\si' }
          + \sum_{k \in L,R; \si} \left( \Vkt \akd \cs +
                                           \mbox{h.c.}  \right).  
\eneq
Here $ \akd (\ak) $ creates (annihilates) an electron with momentum
 $ k $ and spin $ \si $ in the left ($k \in L$) or the right
($k \in R$) lead, $ \csd (\cs) $ creates (annihilates) an electron
 with spin $ \si $ in the dot, 
$ \ek $ and $ \ed $ are single - particle energies in the leads and 
the dot respectively, $ U $ is the Coulomb interaction energy in the dot 
and $ n_{\si} \equiv \csd \cs $. 
The transfer matrix elements $ \Vkt $ 
between the leads and the dot are assumed 
to be small compared with $\ed$ and $U$. 
The external fields are included 
through potential shifts of the leads $ \Delta_{L(R)}(t) $, 
defined as, 
\beq                                    \label{Delta}
  \Delta_{L(R)}(t)  \equiv  \phi_{L(R)} + 
                               W_{L(R)} \cos \left( \Omega t +
\alpha_{L(R)}
 \right). 
\eneq
The first term above 
describes a constant potential bias, while the second one is 
due to an alternating field, which, for simplicity, is assumed to be 
monochromatic. We note that the
chemical potentials in the leads are shifted 
by the same amount $ \Delta_{L(R)}(t) $ as the single-particle energies, 
hence the population of energy levels 
in the leads remains intact. 

As far as the value of the parameters is concerned, 
our attention here is focused on the Kondo regime 
which is determined by the conditions, 
\beq                                         \label{KondoRegime}
   \ed < 0, \hspace{1.0cm} \ed + U > 0, \hspace{1.0cm} 
     \left| \ed \right|,  \ed + U  > \Gamma_{\si}, 
\eneq
where 
$ \Gamma_{\si} =  2 \pi  \sum_{k \in L,R} \left| \Vkt \right|^{2}
\delta (\ed - \ek) $ are the widths of the energy levels in the dot. 
Furthermore, it is assumed that the external fields are not strong 
enough to draw the system out of this regime, so that: 
\begin{mathletters}
\label{ExtFcond}
\beq                                     \label{ExtFcond_dc}
    \left| \phi_{L(R)} \right| < \left| \ed \right|, \ed +  U  \text{ and }
\eneq
\beq
    \Omega, W_{L},W_{R} < \left| \ed \right|, \ed + U.  
                                           \label{ExtFcond_ac}
\eneq
\end{mathletters}
It should be stressed however that these conditions 
do not imply a linear response regime.
The latter is defined by the conditions 
$ \left| \phi_{L(R)} \right|, W_{L(R)} \ll T $ while 
$ T \ll \left| \ed \right|,  \ed + U   $. 

At this point it is convenient to apply a canonical transformation 
\cite{BruderSchoeller94,IngoldNaz_SCT92} on the Anderson model 
(\ref{AndersonModel}) whose purpose is to transfer the dependence on 
time into that part 
which contains a small parameter. The transformation is 
defined as follows: 
\bea						\label{CanTransf} 
   H_{A} & = & {\cal U}^{-1} \tilde{H}_{A}{\cal U}  -  
		      i {\cal U}^{-1} \frac{\partial}{\partial t} {\cal U} \\
  {\cal U}(t) & = & \exp \left\{ \frac{i}{\hbar} 
	\int_{-\infty}^{t} \!\! d \ta \left[ \Delta_{L}(\ta) N_{L} + 
	\Delta_{R}(\ta) N_{R} \right]  \right\}.   \nonumber  
\enea 
where $ \hat{N}_{L(R)} \equiv \sum_{\si, k \in L(R)} \akd(t) \ak(t) $. 
The Hamiltonian $H_{A}$ resulting from this transformation reads, 
\bea			                                     \label{AndersonHam}
   H_{A} & = & \sum_{k \in L,R; \si}  \ek  \akd \ak 
             + \sum_{\si} \ed \csd \cs  
             + \frac{1}{2} U \sum_{\si, \sigma' \ne \si} n_{d, \si} n_{d, 
\si' } + \nonumber  \\
         & & + \sum_{k \in L,R; \si} \left( \Vk \akd \cs + \mbox{h.c.}  
\right).  
\enea
where 
\beq					\label{Vk}
  \Vk   =  \Vkt  \exp \left\{ \frac{i}{\hbar} 
	\int_{-\infty}^{t} \!\! d \ta  \Delta_{L(R)}(\ta) \right\}. 
\eneq

Usually, in tunneling systems the barriers have low transparency. 
Therefore it is convenient to consider the tunneling part of the 
Hamiltonian, i.e. \linebreak 
$ \sum_{k \in L,R; \si} \left( \Vk \akd \cs + \mbox{h.c.}  \right) $, as a 
perturbation \cite{CarComNozSJ-1-71}. 
It is well known, however, that if the interaction 
$\frac{1}{2} U \sum_{\si, \sigma' \ne \si} n_{d, \si} n_{d, \si' }$ is 
kept in the ``unperturbed'' Hamiltonian the Wick's theorem can not be 
applied. In order to circumvent this problem we assume that 
the energies associated with the direct and alternating
voltages as well as the pertinent frequencies are 
smaller than the Coulomb interaction energy 
$U$ in the dot. In other words, 
the assumption $U \rightarrow \infty$ 
should be an excellent approximation.
We then apply the method of slave (auxiliary) bosons 
\cite{Coleman84,Abrikosov65,Barnes76,Bickers87} using 
a certain version of it which is due to
Coleman\cite{Coleman84}. Accordingly, the ordinary electron 
operators in the dot $\cs$, $\csd$, which transform a singly occupied 
state into an empty one and vise versa, are factored into a boson 
operator and a fermion operator \cite{MeirWingreenKondo94}: 
\bea                                   \label{SBdef}
  \cs = \bd \fs  \\
  \csd = b \fsd . \nonumber 
\enea
The slave boson operator $\bd$ ($b$) creates (annihilates) an empty 
state, while the slave fermion operator $\fs$ ($\fsd$) annihilates 
(creates) a singly occupied state. 
In this representation the Hamiltonian (\ref{AndersonHam})becomes,  
\beq                                      				\label{SBHam}
   H_{SB} =  \sum_{k \in L,R; \si}  \ek  \akd \ak 
             + \sum_{\si} \ed \fsd \fs  
         	 + \sum_{\si, k \in L,R} 
		\left( \Vk \akd \bd \fs + \mbox{h.c.} \right). 
\eneq
The Coulomb interaction does not appear in the Hamiltonian any more. 
Indeed, when $U$ is infinite it 
completely eliminates the possibility of double 
occupancy of the dot. This projection is 
accomplished by including the annihilation 
operator $b$ in equations (\ref{SBdef}), which prevents creation of 
the doubly occupied state, and the constraint that the total number of 
slave bosons and slave fermions, 
\beq                               				\label{Qdef_And}
  Q_{A} \equiv  \bd b  +  \sum_{\si} \fsd \fs , 
\eneq
must be equal to unity. The operator $ Q_{A} $ commutes with the 
Hamiltonian (\ref{SBHam}) so that the ``charge'' $ Q_{A} $ is not 
changed during the evolution of the system. Therefore the requirement $ 
Q_{A} = 1 $ does not really constrain the dynamics of the system.
Rather, it assures that the initial state does not contain 
a doubly occupied state as a 
component. In order to enforce the condition $ Q_{A} = 1 $ Coleman 
\cite{Coleman84} introduced a Lagrange multiplier (chemical potential) 
$ -\lam $ and considered the Hamiltonian $ H_{SB} + \lam  Q_{A} $. 
The calculations are to be done with finite $\lam$ and then, 
at the end, $ \lam \rightarrow \infty $. If the tunneling part of 
$ H_{SB} + \lam  Q_{A} $ is chosen as perturbation the remaining 
``unperturbed'' Hamiltonian is now 
quadratic in the creation and annihilation 
operators. The Wick's theorem can then be applied and
diagrammatic expansion is feasible. We notice that the 
unperturbed part has become a simple time - independent free - particle 
Hamiltonian while the perturbation (tunneling part) contains both 
interaction and time - dependence. 

\subsection{The tunneling current}
The tunneling current from the left (right) lead into the central region is 
defined as the product of electron charge ($-e$) and the rate of
change in the number of electrons in that lead. The latter is obtained
by commutating the number - of - electrons (Heisenberg) operator 
$ \hat{N}_{L(R)} $ with the Hamiltonian (\ref{SBHam}). This yields 
\beq                                         \label{Curr-df_And} 
 I_{L(R)}^{T}(t)  =  - e \left\langle \frac{d\hat{N}_{L(R)}(t)}{dt} 
                                          \right\rangle 
	       =  - \frac{i e}{\hbar} \sum_{\si, k \in L(R)} \Vk 
     \mbox{Tr}_{F_{1}^{A}} \left\{  \rho_{F_{1}^{A}}(0) 
                                             \akd(t) \bd(t) \fs(t)  \right\} 
                + c.c., 
\eneq
where $ \rho_{F_{1}^{A}}(0) $ is the density matrix of the system at 
a certain fixed time which is taken here at 
$ t = 0 $. Generally, it should include all the changes that the 
system has undergone since the tunneling was switched on 
\cite{Keldysh65}. 
We recall that the above equation is written in the Heisenberg
representation. The angle brackets on the left hand side mean, of course, 
quantum average over the physical Hilbert space. At infinite $U$ it is 
only a subspace $F_{1}^{A}$ of the full Hilbert space for the slave - 
boson Hamiltonian (\ref{SBHam}) which is determined by the 
condition $ Q_{A} =1$. It is explicitly 
manifested on the right - hand side of 
the above expression by the subscript $F_{1}^{A}$ which restricts the 
trace to the physical subspace $F_{1}^{A}$. In the following we 
calculate directly the current 
(\ref{Curr-df_And}) without 
prior elaboration on the dot Green 
functions as it was done in Refs. 
\onlinecite{Hershfield91,MWLKondo93,Ng93,%
MeirWingreenKondo94,HettlerSchoeller95,SivanWingreen96,Ng96}. 
Unlike the equation for the {\em DC} \cite{MW92} which 
expresses it in terms of the interacting density of states in the dot, an 
expression for the {\em AC} in terms of
the dot  Green functions 
\cite{WJM93ac,JWM94ac} involves two Green functions ($G^{<}$ 
and $G^{r}$) and integration over real time. We find it easier to 
establish a 
perturbation expansion for the current itself (rather than 
for the Green function). To this end we rewrite 
equation (\ref{Curr-df_And}) in the 
interaction representation using the ``grand - canonical'' Hamiltonian 
$ H_{SB} + \lam Q_{A} $. The result is, 
\beq                                                    \label{Curr_IntRep_And} 
 I_{L(R)}^{T}(t)  = - \frac{i e}{\hbar} \sum_{\si, k \in L(R)}
		\Vk  \mbox{Tr}_{F_{1}^{A}} \left\{  
\rho_{F_{1}^{A}}(0) 
	   \SMr \hat{T} [ \akd(t+0) \bd(t) \fs(t) \SM ] \right\}  +  c.c., 
\eneq
where $ \hat{T} $ is the usual time - ordering operator, 
$ \SM = \hat{T} e^{ -i \int_{-\infty}^{+\infty} \! \! H_{T} dt  }$ is the 
usual S - matrix and $ \SMr = S^{\dagger}(+\infty,-\infty) $, while 
$ H_{T} \equiv \sum_{k \in L,R; \si} \left( \Vk \akd \cs + \mbox{h.c}  
\right) $ is the tunneling part of the Hamiltonian 
$ H_{SB} + \lam Q_{A} $ which is chosen as a perturbation. The 
operators $ \akd(t+0) $, $ \bd(t) $ and $ \fs(t) $ in the above equation 
now appear in the interaction representation 
(this is self evident as indicated by the 
presence of the $ S $ - matrix). We notice that it does not really matter 
whether one defines the interaction 
representation using the Hamiltonian $ 
H_{SB} + \lam Q_{A} $ or $ H_{SB} $ because $ Q_{A} $ 
commutes with all parts of $ H_{SB} $. The factor $ \SMr $ in equation 
(\ref{Curr_IntRep_And}) prevents derivation of Wick's theorem and 
subsequent development of the Feynman diagrams technique. Indeed, 
the derivation of Wick's theorem is based on commutation of operators 
which produces pairings, namely ``contractions'' (see Ref. 
\onlinecite{FetterWalecka_book71}). Commutation of two operators 
can not be 
worked out if one of them is subject to the time - ordering operator 
while the other one is not. The idea of a closed time path was 
introduced \cite{Schwinger61,KadBaym62,Keldysh65} in order to treat the 
troublesome factor $ \SMr $ in non-equilibrium systems. The normal 
time branch is continued and turned back so that 
this operator becomes a factor within the scope of 
the time - ordering operator ($ \Tp $) on the closed time - path. 
Equation (\ref{Curr_IntRep_And}), then, can be expressed as,   
\beq                                                    \label{Curr-CTPath_And} 
 I_{L(R)}^{T}(t)  = - \frac{i e}{\hbar} \sum_{\si, k \in L(R)}
		\Vk  \mbox{Tr}_{F_{1}^{A}} \left\{  
\rho_{F_{1}^{A}}(0) 
			    \Tp [ \akd(t_{+}) \bd(t_{-}) \fs(t_{-}) \SMp ] 
\right\}  +    c.c.  
\eneq
Here $ \SMp = e^{  -i \int_{p} \! \! H_{T} dt_{p}  } $, where $ t_{p} 
$ is the variable on the closed time - path and $ \int_{p} $ means 
integration over it. The subscript ``$ - $'' on $t$
signifies that the instant of 
time is considered on the normal (forward) time branch, while the 
subscript ``+'' is used for the backward oriented 
time branch (see Fig. 
\ref{Newfig2}). The choice of the subscripts ``$ + $'' and ``$ - 
$'' for the time arguments of
the operators $ \akd(t_{+}) $, $ \bd(t_{-}) $ and $ \fs(t_{-}) $ 
is consistent in order
to assure that their ordering by the operator $ \Tp $ 
is the same as in Eq. (\ref{Curr-df_And}). The procedure of 
transformation from the Heisenberg representation (Eq. 
(\ref{Curr-df_And})) to the closed time path representation (Eq. 
(\ref{Curr-CTPath_And})) is quite familiar 
\cite{Keldysh65,SuYu91,ChouSuHaoYu85}. 
We note, however, that it has usually 
been employed in the definition of non-equilibrium Green functions while 
here it is applied for the current operator. 

Yet, perturbation expansion of an expression for the current in the form 
(\ref{Curr-CTPath_And}) can not be carried out 
because of the constraint on traces
in the subspace $ F_{1}^{A} $. In order to get rid of it we apply the 
Coleman method \cite{Coleman84} which is feasible here since the equation 
(\ref{Curr-CTPath_And}) appears as an ordinary statistical average of 
the operator 
$ O_{A} \equiv \Tp [ \akd(t_{+}) \bd(t_{-}) \fs(t_{-}) \SMp ] $ in the 
subspace $ F_{1}^{A} $. Let us choose the initial distribution 
of the system (before the tunneling was switched on) 
in the full Hilbert space to 
be an equilibrium one (grand - canonical ensemble with the ``chemical 
potential'' $ - \lam $). Then the initial density matrix 
is, 
\begin{equation}
 \rho(-\infty) \equiv e^{ - \beta ( H_{0} + \lam Q_{A} ) } 
/Z_{G}(\lam) , 
\label{rhoin}
\end{equation}
where $ H_{0} \equiv H_{SB} - H_{T} $ is the unperturbed part of 
the Hamiltonian (\ref{SBHam}) and 
$ Z_{G}(\lam)  \equiv  \mbox{Tr} \left\{ e^{ - \beta ( H_{0} + \lam 
Q_{A} ) } \right\} $ is the grand - canonical partition function. 
Now let us consider the expectation values of the operators 
$ O_{A}Q_{A} $ and $ Q_{A} $ in the full Hilbert space: 
\begin{eqnarray*}
   \langle O_{A}Q_{A} \rangle & \equiv & 
	\text{Tr} \left\{ \rho(0) O_{A} Q_{A} \right\}, \\
   \langle  Q_{A}  \rangle & \equiv & 
	\text{Tr} \left\{ \rho(0) Q_{A} \right\}, 
\end{eqnarray*}
where $ \rho(0) $ is the density matrix of the system in the full Hilbert 
space at the zeroth instant of time. When the tunneling is switched on 
the density matrix evolves in time, so that $ \rho(0) $ differs from $ 
\rho(-\infty) $. However, due to the fact that the operator $ Q_{A} 
$ commutes with the Hamiltonian it can still be 
factorized into separate blocks 
for each subspace $ F_{Q}^{A} $ with different number $ Q_{A} $. 
Thus, 
\bea						\label{OQQ}
  \langle O_{A}Q_{A} \rangle & = & 
	\sum_{Q_{A}=0}^{\infty} 
	\mbox{Tr}_{F_{Q}^{A}} 
		\left\{  \rho_{F_{Q}^{A}}(0) O_{A}  \right\} 
	 Q_{A} e^{ - \beta \lam Q_{A} } Z_{Q_{A}}/ Z_{G}(\lam), \\
  \langle  Q_{A}  \rangle & = & 
	\sum_{Q_{A}=0}^{\infty} 
	 Q_{A} e^{ - \beta \lam Q_{A} } Z_{Q_{A}}/ Z_{G}(\lam), 
				\nonumber 
\enea
where 
$ \rho_{F_{Q}^{A}}(0) $ is the density matrix for the subspace $ 
F_{Q}^{A} $ at the zeroth instant of time, while 
$ Z_{Q_{A}} \equiv  \mbox{Tr}_{F_{Q}^{A}} 
\left\{ e^{ - \beta  H_{0} } \right\} $. 
It is easy to see that in the limit $ \lam \rightarrow \infty $ the ratio of 
the two expressions written above becomes the expectation value of the 
operator $ O_{A} $ in the physical subspace $ F_{1}^{A} $: 
\beq						\label{OFone} 
  \langle O_{A} \rangle_{F_{1}^{A}}  \equiv 
 	 \mbox{Tr}_{F_{1}^{A}} \left\{ \rho_{F_{1}^{A}}(0) 
O_{A} \right\} = 
     \lim_{\lam \rightarrow \infty } 
             \frac{ \langle O_{A} Q_{A} \rangle }{\langle Q_{A} \rangle }. 
\eneq
It is clear that the operator $ O_{A} $ has zero expectation value in the 
subspace $ F_{0} $ because there are neither slave fermions nor  slave 
bosons there. Then $ Q_{A} $ can be dropped out of the numerator in 
(\ref{OFone}). This results in our final expression for the current, 
\beq                                                   	 \label{CurrGCE_A} 
 I_{L(R)}^{T}(t)  = - \frac{i e}{\hbar} \sum_{\si, k \in L(R)}
	\Vk  \lim_{\lam \rightarrow \infty } 
    \frac{ \langle \Tp [ \akd(t_{+}) \bd(t_{-}) \fs(t_{-}) \SMp ] \rangle }
	{\langle Q_{A} \rangle }  +    c.c. 
\eneq
In this equation the averages are taken in the full Hilbert space. It is 
amenable for standard field - theoretical calculation techniques. 

\subsection{Diagrammatic expansion of the current}
In the rest of this section we show how to 
work out a perturbation expansion for the current starting from
equation (\ref{CurrGCE_A}) in powers of the tunneling part of 
the Hamiltonian (\ref{SBHam}). In principle, a careful calculation of 
the density matrix $ \rho(0) $ is required for the non-equilibrium 
perturbation expansion. Otherwise, truncation of the full set of 
diagrams to a finite number might lead to completely wrong results. 
Indeed, truncation of a perturbation set implies that the effect of the 
perturbation is small, i.e. higher order
terms of the perturbation series are negligible. It is not 
always ensured by the small value of the perturbation parameter. 
Consider, for example, an empty dot connected very weakly to a 
reservoir. The smaller the tunneling between them is, the longer it takes 
to fill the dot, but eventually the dot becomes full. It means that the 
occupation of the dot changes by a finite amount although the coupling 
is infinitesimally small. This ``long - times'' perturbation problem  
appears in the energy representation as an infrared divergence 
\cite{SivanWingreen96}. Fortunately, it has been explicitly shown 
\cite{SivanWingreen96} that in the Kondo limit (determined by the 
conditions (\ref{KondoRegime}) and (\ref{ExtFcond_dc})) the 
calculation of the Kondo - type contribution to the {\em DC} up to 
sixth order in $V$ can be done using slave boson and slave fermion
populations for a disconnected dot. In calculating the  
{\em AC} we also have to impose
the condition (\ref{ExtFcond_ac}). The calculation of 
the denominator in equation (\ref{CurrGCE_A}) is, then, very simple: 
\beq 						\label{denom_And}
  \langle Q_{A} \rangle  =  n_{b} + 2 n_{f}  
	\stackrel{\lam \rightarrow \infty}{ \longrightarrow } 
	\left[ 1 + 2 e^{ - \beta ( \ed - \mu ) } \right] e^{ - \beta \lam}, 
\eneq
where 
\bea						\label{FermiBose}
n_{b} & = & 1 / \left( \exp ( \beta \lam ) - 1 \right)  \  \  \mbox{and}  \\
n_{f} & = & 1 / \left( \exp [ \beta ( \ed - \mu + \lam ) ] +1 \right),   
\nonumber
\enea
are Bose and Fermi functions for the slave particles. We assume here, 
for simplicity, that the number of spin degrees of freedom $ \si $ is 
equal to two. 

In order to study the numerator of equation (\ref{CurrGCE_A}) we start 
from its perturbation expansion in powers of $ H_{T} $ (i.e. in powers 
of $ \Vk $). Then we transform every term of the expansion from the 
closed time - path to the single time branch (see Ref. 
\onlinecite{SuYu91,LandauLif81,Mahan-b-90,RammerSm86}). That 
implies expression of the Green functions as matrices in Keldysh 
space (see equations (\ref{matrices_A}) below). The next step is a 
rotation in this space (see Ref. 
\onlinecite{LandauLif81,ChouSuHaoYu85}) resulting in the so called 
\cite{ChouSuHaoYu85} ``physical representation'' for these matrices. 
Finally, time - translation invariance of the unperturbed Green 
functions allows us to apply Fourier transform and work in the energy 
representation. 

The diagrams of forth and sixth order in $ \Vk $ are shown in 
Fig. \ref{Newfig3}. Recall that disconnected diagrams need not 
be considered in the Schwinger - Keldysh formalism. Contribution of a 
disconnected diagram can be factorized into contributions of the 
connected and disconnected parts. Summation over all possible 
disconnected parts yields $ \left\langle \Tp S_{p} \right\rangle $ that is 
$ \left\langle S(-\infty, -\infty) \right\rangle $ which is equal to unity. 
Notice, that diagram ``h'' is a crossed diagram. 
It is not included in the ever used
non - crossing approximation. 
We now formulate the basic rules for drawing the diagrams and 
writing down the corresponding analytical expressions.

\begin{enumerate}
	\item 
In order to obtain a diagram of the $ m^{\text{th}} $ order in $ V $, 
draw a circle (it appears as a polygon in the figures)
consisting of $ m $ alternating slave - boson and slave - 
fermion lines going in the same direction. 
The number $ m $ has to be even because every tunneling vertex 
contains only one slave - boson and one slave - fermion operator. The 
diagram must include only one such circle, otherwise it would produce 
contribution to the current of first or higher order in powers of 
$ \exp ( - \beta \lam ) $ (see Ref. \onlinecite{Coleman84}) that vanishes 
at $ \lam \rightarrow \infty $. We represent the diagrams by closed 
circles rather than by open lines as it is usually done since the external 
operators $ \bd $ and $ \fs $ in equation (\ref{CurrGCE_A}) are taken at 
the same time $ t_{-} $. Connect the vertices by lead - electron lines. 
Remember that only the two types of vertices shown in Fig. 
\ref{Newfig4} are allowed. 
	\item Introduce the following $ 2 \times 2 $ (Keldysh) matrices 
for every lead - electron, slave - fermion and slave - boson lines 
respectively:
\beq						\label{matrices_A}
  g_{k \si}\epbr \equiv \left( \begin{array}{cc}            
			  0   &   g^{a}_{k \si}\epbr  \\                   
                      g^{r}_{k \si}\epbr &   g^{c}_{k \si}\epbr           
                         \end{array}      \right),  \  \  \  
  \xi_{\si}\ombr  \equiv  \left( \begin{array}{cc}            
			  0   &   \xi^{a}_{\si}\ombr   \\             
                      \xi^{r}_{\si}\ombr &   \xi^{c}_{\si}\ombr           
                         \end{array}      \right), \ 
\eneq
\[
  d(\nu)  \equiv  \left( \begin{array}{cc}            
			  0   &   d^{a}(\nu)   \\       
                      d^{r}(\nu) &   d^{c}(\nu)           
                         \end{array}      \right) . 
\]
The indices ``r'', ``a'' and ``c'' stand to denote retarded, advanced and 
correlation 
Green functions. They are defined explicitly as follows: 
\bea					\label{unperturbedGF_A}
  g^{r(a)}_{k \si}\epbr & = &	
      \frac{1}{\epsilon - \ek + \mu \pm i \gam}  \\
  g^{c}_{k \si}\epbr & = &			   
      \left[ 1-2 f(\ek) \right]  \left[ g^{r}_{k \si}\epbr  - g^{a}_{k 
\si}\epbr \right]   \nonumber \\ 
  \xi^{r(a)}_{\si}\ombr  & = &			
      \frac{1}{\omega - \ed + \mu - \lam \pm i \gam}  \nonumber \\ 
  \xi^{c}_{\si}\ombr  & = &			      
      \left[ 1 -  2 \exp \left( - \beta ( \ed - \mu ) \right) \exp ( -\beta \lam )  
\right]  
	\left[ \xi^{r}_{\si}\ombr  - \xi^{a}_{\si}\ombr  \right] 
\nonumber \\ 
  d^{r(a)}(\nu)  & = &			
      \frac{1}{\nu - \lam \pm i \gam}  \nonumber \\ 
  d^{c}(\nu)  & = &			      
      \left[ 1 -  2  \exp ( -\beta \lam )  \right]  
	\left[ d^{r}(\nu) - d^{a}(\nu)  \right], \nonumber 
\enea 
where $ f(\ek) = 1 / \left( \exp [ \beta ( \ek - \mu ) ] +1 \right) $ 
is the Fermi 
function for lead electrons, the factors 
$ \exp [ - \beta ( \ed - \mu )] \exp ( -\beta \lam ) $ and 
$ \exp ( -\beta \lam ) $ are the 
limiting forms of the Fermi and the Bose 
functions (\ref{FermiBose}) respectively at large $ \lam $. 
We note that different representations can be used for the matrices 
(\ref{matrices_A}) (see Ref. \onlinecite{ChouSuHaoYu85,%
LandauLif81,Mahan-b-90,RammerSm86}). Here we employ the so 
called \cite{ChouSuHaoYu85} ``physical representation''. 
	\item Introduce the following tensors for every internal vertex 
(represented by a closed circle in Fig. \ref{Newfig3}). For the vertex 
drawn in Fig. \ref{Newfig4}a:
\beq 
  \Vkt  J_{s}\left( W_{(k)}/ \Omega \right)  	\label{dotlead} 
	e^{ i s \alpha_{(k)}}  \eta^{m}_{i j} 
\eneq
For the vertex drawn in Fig. \ref{Newfig4}b:
\beq 
  \Vkt^{*}  J_{q}\left( W_{(k)}/ \Omega \right)     \label{leaddot} 
	e^{ - i q \alpha_{(k)}}  \eta^{m}_{i j},  
\eneq
where $ J_{s} $ and $ J_{q} $ are Bessel functions. The index ``$ (k) 
$'' reminds us that, despite the fact 
that $ W $ and $ \alpha $  do not depend 
on $ k $, they depend on  the lead to which $ k $ belongs. 
The factors 
$ J_{s(q)}( W_{(k)}/ \Omega ) \exp ( \pm i s(q) \alpha_{(k)} ) $  
originate from the time - dependence of $ \Vk $. Indeed, substituting 
(\ref{Delta}) into (\ref{Vk}) we find 
\bea		
  \Vk  & = & \Vkt  e^{ \frac{i}{\hbar} \left[
     \phi_{(k)} t  +  i  W_{(k)}/ \Omega 
        \sin \left( \Omega t + \alpha_{(k)} \right) -  \beta_{(k)} \right]  } \\
          & = & \Vkt  \sum_{s=-\infty}^{\infty} J_{s}\left( W_{(k)}/ 
\Omega \right) 
	e^{ \frac{i}{\hbar} 
	       \left[ \phi_{(k)} t + s \Omega t + s \alpha_{(k)} - 
\beta_{(k)} \right] }   	
						\nonumber
\enea
The time - dependent factors appearing in this 
expression will surface when we impose  
energy conservation (see below), the phases $ \beta_{(k)} $ cancel, 
while the rest enters equations (\ref{dotlead}) and (\ref{leaddot}). 
The physical meaning of 
$ s $ is the number of photons emitted when an electron goes from the 
dot to a lead, while  $ q $ is the number of photons absorbed by an 
electron going from a lead to the dot. 
Both $ s $ and $ q$ can assume negative values.
The tensor $ \eta $ is given by the following expressions:
\beq						\label{eta}
  \eta^{1}_{i j} =  \frac{1}{\sqrt {2}}  
		       \left( \begin{array}{cc}            
				  1  &  0  \\                   
		                          0  &  1          
                         		\end{array}      \right),  \  \  \ 
  \eta^{2}_{i j} =  \frac{1}{\sqrt {2}}  
		       \left( \begin{array}{cc}            
				  0  &  1  \\                   
		                          1  &  0          
                         		\end{array}      \right).    
\eneq
It appears as a result of the transformation leading from
from time integration over past-ward going time
branch to integration over normal time axis which requires a change of 
sign. The above  form is consistent with the physical 
representation for the Keldysh matrices and tensors \cite{Keldysh65}. 
\\ 
For the external vertex (represented by an open circle in Fig. 
\ref{Newfig3}) containing the operators 
$ \akd(t_{+}) $, $ \bd(t_{-}) $ and $ \fs(t_{-}) $ of equation 
(\ref{CurrGCE_A}) write the factor 
$ J_{s}( W_{L}/ \Omega ) \exp ( i s \alpha_{L} ) $ as for the internal 
vertices but do not insert the tensor $ \eta $. Instead, close around the 
product of  matrices and tensors introduced above by the row 
$ 2^{-1/2} ( 1, 1)  $ at the $ \xi $ - matrix and the vectors 
$ \frac{1}{\sqrt {2}} 
\left( \begin{array}{c}  -1 \\  1  \end{array} \right)  $  and 
$ \frac{1}{\sqrt {2}} 
\left( \begin{array}{c}   1 \\  1  \end{array} \right)  $ 
at the matrices $ g $ and $ d $ respectively as we show in the example 
(equation (\ref{diaE_A})) below. The external vertex differs from the 
internal ones because the time - variable for every external  operator 
$ \akd(t_{+}) $, $ \bd(t_{-}) $ or $ \fs(t_{-}) $ is chosen on a certain 
($ + $ or $ - $) branch and these time
 branches are different for $ \akd $, $ \bd 
$ and $ \fs $. In an internal vertex, the time - variables for all the 
operators are chosen on a single time
branch following by summation over the 
two time branches. 
	\item Conserve spin in every vertex. 
As for energy, it should be conserved in every 
internal vertex taking into account emission (absorption) of $ s $ ($ q $) 
photons and the different values assumed by
the static potential energy $ \phi_{(k)} $ 
on the left and right lead. 
The energy in the external vertex is not conserved. 
The pertinent energy difference 
is equal to the frequency of the current. 
Therefore, introduce for this vertex the factor 
$ \delta _{ n, (q - s + q_{1} - s_{1} + \ldots ) } $, where $ n $ has the 
meaning of the total number of absorbed photons. Summation over $ n 
$ will be carried out in equation (\ref{CurrRuled}) below. 
	\item Sum over energy, momentum and spin. Do not specify to 
which lead the momenta belong. It will be taken care 
of later on. 
	\item Sum over numbers of photons $ q, s, q_{1}, \ldots $. 
	\item Multiply the result  by the factor 
\beq 
  \frac{ - i e}{ \hbar \langle Q_{A} \rangle }  (-i)^{m-1} 
	 (i)^{3m/2}  (-1)^{F}  (2\pi )^{-m/2} 
\eneq  
where $ F $ is the number of closed electron - fermion loops. 
The first factor in the above 
expression comes from equation (\ref{CurrGCE_A}), the second 
one is implied by 
the $ m - 1 $ order of the expansion of the $ S $ - matrix. 
As for the next two factors, recall that an
application of the Wick's theorem 
results in pairings of creation and
annihilation operators which are, then, expressed through  Green 
functions. The 
last factor simply results from the Fourier transform. 
\item At this stage 
take the limit $ \lam \rightarrow \infty $.
	\item Let us denote the expression obtained through the rules 
that have been listed by the symbol $ \Upsilon_{A}(n) $. Then the 
current is given by the following equation: 
\bea					\label{CurrRuled}
  I_{L(R)}^{T}(t) & = & \frac{1}{\pi} \sum_{n} \sum_{ 
(k_{1}),(k_{2}),\ldots =L,R } 
	\left\{ \mbox{Re} \left[ \Upsilon_{A}(n) \right]  
	\cos \left( n \Omega t - \alpha_{(k),(k_{1}),\ldots} \right)     
\right.   \\ 
   & & \hspace{4cm}   + \left. \mbox{Im} \left[ \Upsilon_{A}(n) \right]  
	\sin \left( n \Omega t - \alpha_{(k),(k_{1}),\ldots} \right) 
\right\},   \nonumber 
\enea
where 
$ \alpha_{(k),(k_{1}),\ldots} \equiv  (s-q) \alpha_{L} + 
    (s_{1}-q_{1}) \alpha_{(k_{1})} +  (s_{2}-q_{2}) 
\alpha_{(k_{2})} +  \ldots $, while 
$ \sum_{ (k_{1}),(k_{2}),\ldots =L,R } $ means summation over the 
leads to which $ k_{1} $, $ k_{2} $, etc. belong. There is no summation 
over $ (k) $. The electron line coming out of the external vertex has 
momentum $k$ belonging only to the left (right) lead since  we are 
calculating current through the left (right) barrier. The other momenta 
run through both leads.  
\end{enumerate} 

As an example we write below
 the expression which is obtained using the 
rules listed above for the diagram drawn in Fig. \ref{Newfig3}e: 
\bea                                \label{diaE_A}
\Upsilon_{A}(n) & = & \lim_{\lam \rightarrow \infty }
	\frac{ - i e}{\hbar \langle Q_{A} \rangle } 
	\frac{1}{16} \frac{1}{(2\pi )^{3}}  
	\sum_{q, \qa, \qb, s, \ssa, \ssb} 
	\delta _{ n, (q + \qa + \qb - s - \ssa - \ssb ) }       \\
& \cdot  &	J_{q}\left( \frac{W_{L}}{\Omega} \right) 
	J_{s}\left( \frac{W_{L}}{\Omega} \right)
	J_{\qa}\left( \frac{W_{(\ka)}}{\Omega} \right)
	J_{\ssa}\left( \frac{W_{(\ka)}}{\Omega} \right)
	J_{\qb}\left( \frac{W_{(\kb)}}{\Omega} \right) 
	J_{\ssb}\left( \frac{W_{(\kb)}}{\Omega} \right)  
                                                           \nonumber   \\
& \cdot  &	\sum_{\ka \in L; k, \kb \in L,R; \si, \sib} 
	\left| \Vkt \right|^{2} \left| \Vkta \right|^{2} 
                                             \left| \Vktb \right|^{2} 
	\int \!\! \int \!\! \int \!\! \int d \oma \, d\omb \, d\nua \, d\nub \,  
                                                              \nonumber   \\ 
& \cdot  &	\sum_{i, j, m, \ldots, \ie, \je, \me}  (1, \ 1)_{i} 
	\xi^{i \ia}(\oma)  \eta^{\ma}_{\ia \ja} 
	g_{\ka \si}^{\ja \jb} (\oma - \nu -\phi_{(\ka)} -\qa \Omega) 
            d^{\ma \mb}(\nu)  \eta^{\mb}_{\ib \jb}   \nonumber \\
& \cdot & \xi^{\ia \ib}(\oma - (\qa-\ssa)\Omega) \eta^{\mc}_{\ic \jc} 
	g_{k \si}^{\jc j} 
	(\oma - \nua -\phi_{L} - (q + \qa + \qb-\ssa-\ssb) \Omega) 
                                                              \nonumber   \\
& \cdot &	d^{\mc \md}(\nua + (\qb-\ssb)\Omega)                   
	\eta^{\md}_{\id \jd} \xi^{\id \ie}(\omb ) \eta^{\me}_{\ie \je} 
	g_{\kb \sib}^{\je \jd} (\omb - \nua -\phi_{(\kb)} - \qb \Omega) 
                                                             \nonumber   \\
& \cdot &	d^{\me m}(\nua ) 
	\left( \begin{array}{c}  -1 \\  1  \end{array} \right)_{j} 
	\left( \begin{array}{c}   1 \\  1  \end{array} \right)_{m}. 
\nonumber 
\enea

In the above equation the tensor product is to be expanded followed by
integration over internal variables $ \oma, \, \omb, \, \nua $ and $
\nub $ is to be carried out. 
Using the {\em Mathematica} program we 
have carried it out for the 
above equation and for the analogous one given by 
the diagram drawn in Fig. \ref{Newfig3}f. Whereas 
the calculation of the current could not be completed 
(see below) it was possible to inspect the emergence of 
a  Kondo behavior through the appearance of a term 
$\sum_{p} \frac{f(p)} {\epsilon_{k}-\epsilon_{p}}$ where 
$f(p)=f(\epsilon_{p})$ is the Fermi function for lead 
electrons of quantum number $p$. It is well known that 
this term is characteristic for the Kondo effect 
(above the Kondo temperature). 
Accordingly, the resulting expression 
for diagram \ref{Newfig3}f does not exhibit a Kondo - type 
behavior (i.e. $ \ln T $ - divergence in the linear response) while the 
one obtained for diagram \ref{Newfig3}e and 
represented by equation (\ref{diaE_A}) does. 
The equations in their final form 
are very long and cumbersome and will not be
shown here. Moreover, expressions appearing in 
the intermediate stages are unusually long and 
often overflowing the memory of 
a typical present - day workstation. 
We stress that this occurs only 
when computing the {\em AC} (calculations appropriate for 
the {\em DC} are much simpler). These manipulations 
pertaining to {\em AC}
appear to be especially cumbersome for the crossed diagram 
\ref{Newfig3}h. 
We argue that, physically, 
the main source of complications results from the 
admixture of the Kondo resonance with usual resonant tunneling 
which is present in the Anderson model. Although the latter 
contribution is exponentially 
small within the relevant range of parameters, governed by terms like
$ \exp \left( \ed / (k T ) \right) $ and may frequently be 
neglected in the Kondo regime, its omission
can not be assumed in this calculation scheme before 
arriving at the final stages. We are therefore content 
by having introduced the systematic calculation scheme for 
the general time - dependent Anderson model, and 
exposing some diagrams that manifest a Kondo 
behavior. With rapid improvement of 
workstation capacities and analytic software programs 
the evaluation of all diagrams of sixth order presented here 
can be completed according to the rules specified above. 
With this final note we now move on 
to complete the calculation of the current within an 
effective theory which still captures most of the 
pertinent subtle physics.  

\section{Time - dependent Schrieffer - Wolff transformation}
\label{TDSW}

In this section we transform the time - dependent Anderson Hamiltonian 
to a time - dependent Kondo Hamiltonian. We use a procedure similar to the 
one proposed by Schrieffer and Wolff \cite{SchriefferWolff66} for the 
time - independent Anderson model but extend it to time - dependent 
problems. 

As we did in the previous section we start from the canonical 
transformation (\ref{CanTransf}) of the Anderson Hamiltonian 
(\ref{AndersonModel}) whose purpose is to transfer the dependence on 
time into the perturbation part. It results in the Hamiltonian 
$ H_{A} $ (Eq. (\ref{AndersonHam})). 
Then we further transform this Hamiltonian as, 
\beq						\label{SWtransf1} 
   H_{A}'  =  e^{S} H_{A} e^{-S}  - 
		      i e^{S} \frac{\partial }{\partial t} e^{-S}, 
\eneq 
where the operator $ S $ is defined 
according to a specific prescription. In the spirit of Ref. 
\onlinecite{SchriefferWolff66} we expand the right hand side of Eq. 
(\ref{SWtransf1}) in powers of $ S $. We notice that $ S $ does not 
commute with $ \frac{\partial S}{\partial t} $ 
so this manipulation should be done with care.
Then we require the operator $ S $ to satisfy the 
equation, 
\beq                                            \label{Seq}
   i \frac{\partial S}{\partial t} + H_{T} + [ S, H_{0} ] = 0. 
\eneq
Here $ H_{T} \equiv \sum_{k \in L,R; \si} \left( \Vk \akd \cs + 
\mbox{h.c}  \right) $ 
is the tunneling part of the Hamiltonian $ H_{A} $ and 
$ H_{0} \equiv \sum_{k \in L,R; \si}  \ek  \akd \ak 
             + \sum_{\si} \ed \csd \cs  
             + \frac{1}{2} U \sum_{\si, \sigma' \ne \si} n_{d, \si} n_{d, 
\si' } $ 
is the rest of it. If  $ S $ is proportional to $ V $ (which is indeed the 
case as we show at the end of this section) 
then solving Eq. (\ref{Seq}) eliminates 
from the Hamiltonian $ H_{A}' $ terms which are of first order in $ V $, 
leaving only terms of higher orders which are naturally smaller. Upon 
collecting terms of zero and second order in $ V $ the 
Hamiltonian $ H_{A}' $ can be written in the following form: 
\beq                                            \label{SWtransf2}
   H_{A}' =  H_{0} + \frac{1}{2} [ S, H_{T} ]  
\eneq
Following Ref.\cite{SchriefferWolff66} we omit terms of 3d and higher 
orders in $ V $. Our transformation differs from that of 
Ref.\cite{SchriefferWolff66} due to the presence of the term 
$ - i e^{S} \frac{\partial }{\partial t} e^{-S} $ in Eq. 
(\ref{SWtransf1}) and, correspondingly, the term 
$ i \frac{\partial S}{\partial t} $ in Eq. (\ref{Seq}). 

Now we turn to the task of solving Eq. (\ref{Seq}), 
and look for a solution of the form, 
\beq                                   \label{Sform}
   S = S_{1} - S_{1}^{\dagger}.
\eneq
The operator $ S_{1} $, then, should satisfy the equation
\beq                                            \label{S1eq}
   i \frac{\partial S_{1}}{\partial t} + H_{T}^{out} + [ S_{1}, H_{0} 
] = 0,
\eneq
where $ H_{T}^{out} \equiv \sum_{k \in L,R; \si} \Vk \akd \cs $ is the 
part of the Hamiltonian responsible for tunneling out of the dot. The 
operator $ S $ is anti - hermitian which assures 
the hermiticity of $ H_{A}' $.

We are now looking for a solution of equation (\ref{S1eq}) in the 
form, 
\beq                                            \label{S1form}
   S_{1} = \sum_{k \in L,R; \si} \Ak 
	\left[ \frac{n_{d, -\si}}{\ek-\ed-U} + \frac{(1-n_{d, 
-\si})}{\ek-\ed} \right] 
	\akd \cs.
\eneq
The sum appearing in the square brackets is the inverse of the operator 
$ \zeta_{k, \si} \equiv \ek-\ed-U n_{d, -\si} $. Schrieffer and Wolff 
found $ \Ak = \Vkt $, where $ \Vkt $ is, of course, time - 
independent. 

Substituting expression (\ref{S1form}) into equation (\ref{S1eq}) we 
find that $ \Ak $ must satisfy the following equation:
\beq                                            \label{Aeq}
   i \dot{A}_{k}(t) \zeta_{k, \si}^{-1} - \Ak + \Vk = 0. 
\eneq
A proper solution of this equation is 
\beq                                                \label{Asol}
   \Ak = {\cal V}_{kd} 
	\sum _{s = -\infty}^{\infty} J_{s}(W_{(k)}/\Omega) 
	e^{ i (\phi_{(k)} + s \Omega ) t + i s \alpha_{(k)} }
	\frac{\zeta_{k, \si}}{ \phi_{(k)} + s \Omega + \zeta_{k, \si} }, 
\eneq
where 
$ {\cal V}_{kd} \equiv \tilde{V}_{kd} 
            \exp \left[ - i \left(W_{(k)} / \Omega \right) \sin \alpha_{(k)}
                        \right]  $.
Recall that the symbol $ (k) $ means ``L'' or ``R'' 
depending on whether $ k $ 
belongs to the left or to the right lead, while 
$ J_{s}(\frac{W}{\Omega}) $ are Bessel's functions. 
The general solution of Eq. (\ref{Aeq}) contains also a
term: $ \propto e^{-i \zeta_{k, \si} t} $, but the 
requirement that $ \Ak $ should be time - independent in the absence of 
external potentials enforces the prefactor to vanish. 

Employing now equations (\ref{SWtransf2}), (\ref{Sform}), 
(\ref{S1form}) and (\ref{Asol}) we obtain the desired form of
the Hamiltonian, in exact 
correspondence with Ref.\cite{SchriefferWolff66},
\bea                                     \label{SchWfnl}
   H_{A}' & = & H_{0} + \sum_{k, k' \in L,R; \si} \! 
	\left( \Wkpk - 1/2 \Jkpk n_{d, -\si} \right) a_{k', \si}^{\dagger} 
\ak   \\  
& & + 1/2 \!\!  \sum_{k, k' \in L,R; \si} \! \Jkpk a_{k', -\si}^{\dagger} 
	\ak \csd c_{d, -\si}     \nonumber   \\
&  & - \sum_{k \in L,R; \si} \! 
	\left( W_{kk}(t) - 1/2 J_{kk}(t) n_{d, -\si} \right) \csd \cs   
\nonumber   \\ 
& & - 1/4 \!\!  \sum_{k, k' \in L,R; \si} \! 
	\left[ \Jkpk a_{k', -\si}^{\dagger} \akd \cs c_{d, -\si}  
		+  h.c. \right],      \nonumber   
\enea
where 
\bea                                     \label{Jkpk}
  \Jkpk &  = & {\cal V}_{k'd} {\cal V}_{kd}^{*} 
    \exp \left[ i \left( \phi_{(k')} - \phi_{(k)} \right) t \right]  \cdot \\
    & & \cdot 
    \sum_{s', s = -\infty}^{+\infty} 
   J_{s'}(\frac{W_{(k')}}{\Omega}) J_{s}(\frac{W_{(k)}}{\Omega}) 
    \exp \left[ i (s' -s) \Omega t + 
                   i (s' \alpha_{(k')} -s \alpha_{(k)}) \right]
                                      \cdot  \nonumber \\
    & & \cdot \left( 
       \frac{1}{\epsilon_{k'} + \phi_{(k')} + s' \Omega - \ed}+ 
       \frac{1}{\ek + \phi_{(k)} + s \Omega - \ed} + \right. \nonumber  \\
    & & \left.  \text{       } + 
       \frac{1}{\epsilon_{k'} + \phi_{(k')} + s' \Omega - \ed - U}+ 
       \frac{1}{\ek + \phi_{(k)} + s \Omega - \ed - U}   \right), 
                                                     \nonumber  \\
  \Wkpk &  = & \frac{1}{2} {\cal V}_{k'd} {\cal V}_{kd}^{*} 
    \exp \left[ i \left( \phi_{(k')} - \phi_{(k)} \right) t \right] 
                                      \cdot  \nonumber \\
    & & \cdot 
    \sum_{s', s = -\infty}^{+\infty} 
   J_{s'}(\frac{W_{(k')}}{\Omega}) J_{s}(\frac{W_{(k)}}{\Omega}) 
    \exp \left[ i (s' -s) \Omega t + 
                   i (s' \alpha_{(k')} -s \alpha_{(k)}) \right]
                                      \cdot     \nonumber  \\
    & & \cdot \left( 
              \frac{1}{\epsilon_{k'} + \phi_{(k')} + s' \Omega - \ed}+ 
              \frac{1}{\ek + \phi_{(k)} + s \Omega - \ed} 
                                  \right).  \nonumber 
\enea
The form of this Hamiltonian is the same as that 
of Ref.\cite{SchriefferWolff66}
but the matrix elements $ \Jkpk $ and $ \Wkpk $ are evidently distinct. 

In the Kondo regime, the important subspace $ F_{1}^{K} $ of the 
full Hilbert space is that one 
for which the dot is occupied by one electron. 
The last term in the Hamiltonian (\ref{SchWfnl}) is not relevant for this 
subspace. The third term in equation (\ref{SchWfnl}) can be absorbed 
into the definitions of $ \ed $ and $ U $. Besides, we have
$ n_{d, \si} n_{d, -\si} = 0 $ and $ \sum_{\si} n_{d, \si} = 1 $ 
in the subspace $ F_{1}^{K} $, so that the Coulomb interaction term 
(present in $ H_{0} $) vanishes and the one - particle energy term for 
the dot becomes a $c$ - number. The remaining terms represent the 
Kondo (also called ``s - d'') Hamiltonian plus a 
potential scattering term,
\bea                                     \label{KondoHam}
   H_{K} & = & \sum_{k \in L,R; \si} \ek \akd \ak 
	+ \sum_{k, k' \in L,R; \si} \!\!  
	\left( \Wkpk - 1/2 \Jkpk n_{d, -\si} \right) a_{k', \si}^{\dagger} 
\ak + 
						\nonumber  \\
   & & \mbox{} + 1/2 \!\!  \sum_{k, k' \in L,R; \si} \!\!  \Jkpk  
	a_{k', -\si}^{\dagger} \ak \csd c_{d, -\si} . 
\enea
Two comments are in order here:
(i) Any procedure toward calculation of physical quantities should take 
into account the fact that, out of the full Hilbert space, the system is 
projected onto a subspace $ F_{1}^{K} $ in which the dot is occupied 
by one (and only one) electron. 
(ii) At this stage one might be tempted to express the electron creation -  
annihilation operators in the dot through spin operators, thus arriving at 
the familiar form 
\cite{Hewson:book93,Fulde:book91,SchriefferWolff66} of the Kondo 
Hamiltonian. But then one would  realize that the spin operators do not 
obey the usual commutation rules. In order to overcome this obstacle, 
fictitious (auxiliary) fermions might be introduced \cite{Abrikosov65}. 
But this leads one back to equation (\ref{KondoHam}). In other words, 
auxiliary fermions which are sometimes regarded as artificial particles 
introduced to represent spins are real electrons in the dot (impurity 
atom) subject to the constraint specified in (i).

Calculation of the tunneling current starting from the Kondo 
Hamiltonian (\ref{KondoHam}) is possible for arbitrary field strengths 
and frequency provided the inequalities (\ref{ExtFcond}) are satisfied. 
Yet, inspecting a typical experimental setup 
\cite{GoldhaberGordon98,Cronenwett98} one may consider somewhat 
weaker external fields and lower frequencies, so that 
\beq                                             \label{ExtFcond_small}
  \left| \phi_{L(R)} \right|, \Omega, W_{L},W_{R} \ll 
                                              \left| \ed \right|,   \ed + U. 
\eneq
Expressions (\ref{Jkpk}) for $ \Jkpk $ and $ \Wkpk $ then 
significantly simplify. Indeed, at small $ W / \Omega $ the Bessel 
function $ J_{s}(W/\Omega) $ rapidly decreases with increasing $ s $. 
At large $ W/\Omega $ it decays strongly once $ s $ exceeds 
$ W / \Omega $. Therefore we can restrict $ s \Omega $ 
to be less than or of the 
order of $ \text{max}(\Omega, W) $, that is,
$ s \Omega \ll \left| \ed \right|, \ed + U $. 
In the formation of the Kondo resonance the most important 
states are those 
with energies $ \left| \ek \right| \ll \left| \ed \right|, \ed + U $ (see 
Ref. \onlinecite{SchriefferWolff66}). Then 
$ \left|  \zeta_{k, \si} \right| \approx \left| \ed \right| $, 
if $ n_{d, -\si} = 0 $, or 
$ \left|  \zeta_{k, \si} \right| \approx \ed + U $, 
if $ n_{d, -\si} = 1 $ 
(see definition after Eq. \ref{S1form}, 
and recall that $ n_{d, -\si} $ is discrete). Therefore the 
conditions (\ref{ExtFcond_small}) assure that 
\beq
   \phi_{(k)} + s \Omega \ll \left|  \zeta_{k, \si} \right| 
\eneq
(recall that $ \phi_{(k)} $ refers to $ \phi_{L} $ or $ \phi_{R} $). 
Therefore we can neglect the term $ \phi_{(k)} + s \Omega $ in the 
denominator of the expression (\ref{Asol}). Its right hand side 
is greatly simplified, and now becomes,
\beq                                      \label{AeqV}
   \Ak \approx \Vk, 
\eneq
where $ \Vk $ is defined by Eq. (\ref{Vk}). 
Expressions for the
matrix elements $ J_{k'k} $ and $ \Wkpk $ simplify as well. 
First, they become time - independent if $ k $ and $ k' $ belong to 
the same lead. Moreover, they do not depend on potential shifts of 
each lead separately but only on their difference, 
$ \Delta_{L} - \Delta_{R} $. 
For a monochromatic potential difference between the 
leads we then define 
\beq                                          \label{Delta_K}
   \Delta_{LR}  \equiv  \Delta_{L} - \Delta_{R} \equiv 
	\phi^{dc} + W \cos \left( \Omega t + \alpha \right). 
\eneq
Then the matrix elements $ \Jkpk $ and $ \Wkpk $ can be expressed in 
quite a simple form: 
\bea						\label{JW}
   J_{k'k}(t) &  = & \left\{ \begin{array}{ll}  
		\tilde{J}_{k'k} \exp \left[ \frac{i}{\hbar} 
		\int_{-\infty}^{t} \!\! d \ta  \Delta_{LR}(\ta) \right], 
	&	\mbox{if } k' \in L, \; k \in R     \\ 
		\tilde{J}_{k'k} \: ,   
	&	\mbox{if } k', k \in L  \mbox{ or } k', k \in R      
			\end{array}   \right. 
				,   \\
   W_{k'k}(t) &  = & \left\{ \begin{array}{ll}  
		\tilde{W}_{k'k} \exp \left[ \frac{i}{\hbar} 
		\int_{-\infty}^{t} \!\! d \ta  \Delta_{LR}(\ta) \right], 
	&	\mbox{if } k' \in L, \; k \in R     \\ 
		\tilde{W}_{k'k} \: ,   
	&	\mbox{if } k', k \in L  \mbox{ or } k', k \in R      
			\end{array}   \right. 
				,  \nonumber  
\enea
where
\bea						\label{JWtilde}
   \tilde{J}_{k'k} &  \equiv & \tilde{V}_{k'd} \tilde{V}_{kd}^{*} 
        \left( \frac{1}{\ek - \ed} + \frac{1}{\epsilon_{k'} - \ed} -
	\frac{1}{\ek - \ed -U} - \frac{1}{\epsilon_{k'} - \ed - U} 
\right),  \nonumber  \\
   \tilde{W}_{k'k} &  \equiv & \frac{1}{2}  \tilde{V}_{k'd} 
\tilde{V}_{kd}^{*} 
        \left( \frac{1}{\ek - \ed} + \frac{1}{\epsilon_{k'} - \ed}  \right).  
\enea 
are time - independent. The matrix elements $ J_{k'k}(t) $ and 
$ W_{k'k}(t) $ for $ k' \in R, \; k \in L $ satisfy the hermiticity
relations $ J_{kk'}(t)  = J_{k'k}^{*}(t) $ and 
$ W_{kk'}(t)  = W_{k'k}^{*}(t) $. 
We note that at small external fields (\ref{ExtFcond_small}) the 
transformation (determined by Eqs. (\ref{S1form}) and (\ref{AeqV})) 
and the form of the matrix elements $ \Jkpk $ and $ \Wkpk $ are very 
similar to those of Ref.\cite{SchriefferWolff66}
although they are still different 
due to the time - dependence of $ \Vk $. We further note that the 
Hamiltonian (\ref{KondoHam}) with the coupling constants (\ref{JW}) 
can alternatively be obtained by application of the canonical 
transformation (\ref{CanTransf}) to the usual Kondo Hamiltonian 
with time - dependence added only to the leads 
(while the coupling of conduction electrons to the impurity (dot) 
remains time - independent). This is true, however, only at small 
external fields (\ref{ExtFcond_small}). At stronger fields equations
(\ref{Jkpk}) must be used for 
computing the matrix elements $ \Jkpk $ and $ \Wkpk $. 

In order to have a more compact form of the Hamiltonian 
we hereafter imply 
the limit of infinite $ U $. It is obvious from equations 
(\ref{JW}) and (\ref{JWtilde}) that this choice does not lead to any 
qualitative changes. Indeed,
it just slightly affects the values of $ J_{k'k} $. 
The main advantage of this choice is that at $ U = \infty $ the
equality, $ W_{k'k} = 1/2 J_{k'k} $ holds, which eliminates 
one more parameter. 
Furthermore, in the subspace $ F_{1}^{K} $ the following
identity holds, namely,
$ 1- n_{d, -\si} = n_{d, \si} $. Then $ H_{K} $ can 
be expressed as 
\bea                                     \label{KondoFnl}
   H_{K} & = & \sum_{k \in L,R; \si} \ek \akd \ak +   \\
	     &+ & 1/2 \!  \sum_{k, k' \in L,R; \si} \! J_{k'k}(t) 
        \left[ 	a_{k', -\si}^{\dagger} \ak \csd c_{d, -\si} + 
	a_{k', \si}^{\dagger} \ak \csd c_{d, \si}   \right].   \nonumber
\enea
There is only one coupling constant $ \Jkpk $ in this expression which 
is equal for both coupling terms. Equation (\ref{KondoFnl}) constitutes 
our final form of the time - dependent
Hamiltonian which we use in the next section to obtain the 
tunneling current. 

In concluding this section we would like to point out that the main 
idea of the Schrieffer - Wolff transformation is 
based on a projection of the system 
out of the full Hilbert space onto the subspace $ F_{1}^{K} $ for 
which the dot is occupied by one (and only one) electron. Unoccupied 
and doubly - occupied subspaces are forbidden. At first glance 
it looks as a small reduction of space dimension.
Yet, as we show in the next section 
it greatly simplifies the 
calculation of the tunneling current for time - dependent problems. 
The main physical reason seems to be the fact that by
fixing the number of electrons in the dot one separates the Kondo 
resonance from the usual resonant tunneling. The latter is exponentially 
small in the Kondo regime but it is formally present in the Anderson 
model. Admixture of two different physical processes within the same
calculation scheme seems to be the main source of complication. Another 
advantage of the Kondo Hamiltonian (\ref{KondoHam}) (or 
(\ref{KondoFnl})) is readily seen from the expressions 
(\ref{KondoHam}), (\ref{Delta_K}), (\ref{JW}) and (\ref{JWtilde}). 
Namely, the number of independent parameters is significantly reduced. 
Indeed, instead of six parameters controlling the external fields in the 
Anderson model ($ \phi_{L(R)} $, $ W_{L(R)} $, $ \Omega $ and the 
phase difference $ \alpha_{L} - \alpha_{R} $) we are left with only 
three: $ \phi^{dc} $, $ W $ and $ \Omega $. Instead of three independent 
internal parameters $ V_{k d} $, $ \ed $ and $ U $ there are only two 
important combinations $ J_{k' k} $ and $ W_{k' k} $. Moreover, it 
becomes obvious that the limit of infinite $ U $ does not imply any 
qualitative changes in the results. It allows us to get rid of one more 
parameter. 

\section{Perturbation expansion of the current in the Kondo Model}
\label{KondoPerturbation}
\subsection{Expression for the tunneling current}
In this section we define the tunneling current using the time - 
dependent Kondo Hamiltonian (\ref{KondoFnl}) and then develop a 
non-equilibrium technique to expand it in powers of the coupling 
strength $ J_{k' k} $. Although the 
details of calculation are substantially 
distinct from those used in section \ref{Anderson}, 
the basic algorithm is quite similar. 
Of course, we do not need 
to introduce slave particles here, because the unperturbed 
part is bilinear in creation - annihilation operators and there is no 
obstacle in carrying out 
perturbation expansion in powers of the interaction 
which contains a small parameter $ J_{k' k} $. 
The starting point is, again, defining the current using
commutation of the Hamiltonian (\ref{KondoFnl}) with the number - of 
- particles operator. It yields, in the Heisenberg
representation, 
\bea				\label{Curr-df_Kondo}
   I(t) & = & - \frac{i e}{\hbar} 
                    \left\langle \left[ H_{K}, N_{L} \right] \right\rangle 
	 = \frac{i e}{\hbar} 
                \left\langle \left[ H_{K}, N_{R} \right] \right\rangle = \\
	& = & \frac{e}{\hbar} \sum_{k' \in L, k \in R; \si} \! 
        \text{Im} \left\{ J_{k'k}(t) 
	\text{Tr}_{F_{1}^{K}} \left[  \rho_{F_{1}^{K}}(0) 
	a_{k', -\si}^{\dagger}(t) \ak(t) \csd(t) c_{d, -\si}(t)  \right] + 
				\right.  \nonumber \\ 
         & & \hspace{2cm} \left.  \mbox{} + 
            J_{k'k}(t)  \text{Tr}_{F_{1}^{K}} 
            \left[ \rho_{F_{1}^{K}}(0) 
	a_{k', \si}^{\dagger}(t) \ak(t) \csd(t) c_{d, \si}(t)  \right]
					\right\}, \nonumber
\enea
where $ \rho_{F_{1}^{K}}(0) $ is the density matrix of the system at 
the zeroth instant of time (compare with Eq. (\ref{Curr-df_And})). 
As usual, the 
average is taken over the physical subspace $ F_{1}^{K} 
$ of the full Hilbert space for the Kondo model, and the 
subscript $ F_{1}^{K} $ on the right - hand side implies 
that operators and traces are performed 
within this subspace. Unlike the physical 
subspace for the Anderson model (section \ref{Anderson}) which we 
referred to as $ F_{1}^{A} $, the subspace $ F_{1}^{K} $ is defined 
by the condition that $ Q_{K} = 1 $, where 
\beq                               			
	\label{Qdef_Kondo}
  Q_{K} \equiv  \sum_{\si} \csd \cs . 
\eneq
Definition of the physical subspace for the Anderson model, i.e. the 
condition $ Q_{A} = 1 $ (see Eq. (\ref{Qdef_And})) fixed the total 
number of slave particles to be equal to unity, thus, allowing both 
unoccupied and single - occupied states of the dot. The condition 
$ Q_{K} = 1 $, on the other hand,
 enforces single occupation of the dot. 

In order to get rid of the constraint to the subspace $ F_{1}^{K} $ we 
adapt the method proposed by Coleman \cite{Coleman84} for the 
analogous problem in the Anderson model, to be used 
also in the Kondo model. As in section 
\ref{Anderson} we combine it with the Schwinger - Keldysh non - 
equilibrium Green function technique. First, we introduce a grand - 
canonical Hamiltonian $ H_{K} + \lam Q_{K} $. The limit of infinite $ 
\lam $ is to be taken at the end of the calculation. Then we go to the 
interaction representation considering the exchange
interaction as a perturbation and 
the rest of the Hamiltonian, i.e. 
$ \sum_{k \in L,R; \si} \ek \akd \ak + \lam Q_{K} $ as the unperturbed 
part. We notice that $ Q_{K} $ can be freely added to or subtracted 
from the Hamiltonian in the
definition of the interaction representation since 
it commutes with all parts of $ H_{K} $, although it is important in the 
statistical average as we proceed to show below. 
Within the interaction representation, equation
(\ref{Curr-df_Kondo}) for the current now reads,
\bea			\label{Curr-CTPath_Kondo}
   I(t) & = & - \lim_{t' \rightarrow t+0}
	\frac{e}{\hbar} \sum_{k' \in L, k \in R; \si} \! 
        \text{Im} \left\{ \right.           
J_{k'k}(t) 
	\text{Tr}_{F_{1}^{K}} \left[ \rho_{F_{1}^{K}}(0) 
	\Tp ( a_{k', -\si}^{\dagger}(t_{+}) \ak(t'_{+}) 
		\csd(t'_{-})	c_{d, -\si}(t_{-}) S_{p} ) \right] + 
					\nonumber \\ 
         & & \hspace{5mm} \left.   \mbox{} +  
            J_{k'k}(t) \text{Tr}_{F_{1}^{K}} 
	\left[ \rho_{F_{1}^{K}}(0) 
	\Tp ( a_{k', \si}^{\dagger}(t_{+}) \ak(t'_{+}) 
		\csd(t'_{-}) c_{d, \si}(t_{-}) S_{p} ) \right]
				\right\}, \nonumber
\enea
where $ S_{p} $ and $ \Tp $ are, respectively, the S - matrix and the 
time - ordering operator on the closed time - path. In this 
equation and hereafter,
the operators $ \akd(t) $, $ \ak (t) $, $ \csd(t) $ and $ \cs(t) $ 
are defined
in the interaction representation. The choice of points 
$ t_{+} $, $ t_{-} $, $ t'_{+} $ and $ t'_{-} $ on the closed time path 
(see Fig. \ref{Newfig5}) assures proper ordering of these 
operators by the operator $ \Tp $. 
As in section \ref{Anderson} the initial 
distribution of the system in the full Hilbert space 
(before tunneling 
was switched on) is chosen to 
be an equilibrium one. It corresponds to a
grand - canonical ensemble with the ``chemical 
potential'' $ - \lam $. Instead of the operator $ O_{A} $ we consider 
here the operator $ O_{K} $ which is defined by the equation, 
\beq					\label{Odef_K}
   O_{K} \equiv \Tp \left[ \left( 
              a_{k', -\si}^{\dagger}(t_{+}) \ak(t'_{+}) 
                      \csd(t'_{-}) c_{d, -\si}(t_{-})  + 
              a_{k', \si}^{\dagger}(t_{+}) \ak(t'_{+}) 
                      \csd(t'_{-}) c_{d, \si}(t_{-}) 
			\right) S_{p} \right].  
\eneq
Considering expectation values of the operators $ O_{K}Q_{K} 
$ and $ Q_{K} $ within the full Hilbert space and repeating the steps 
leading from Eq. (\ref{Curr-CTPath_And}) to Eq. (\ref{CurrGCE_A}) 
(see equations (\ref{OQQ}) -- (\ref{OFone}) and explanations therein) 
we obtain the following expression for the current: 
\bea					\label{CurrGCE_K}
   I(t) & = & - \frac{e}{\hbar} \sum_{k' \in L, k \in R; \si} \! 
        \mbox{Im} 
	\lim_{t' \rightarrow t+0}  \lim_{\lam \rightarrow \infty } 
	\left\{  \right.     \\
         & & \hspace{5mm} J_{k'k}(t) 
	\frac{ \left\langle 
	\Tp \left[ a_{k', -\si}^{\dagger}(t_{+}) \ak(t'_{+}) 
		\csd(t'_{-})  c_{d, -\si}(t_{-}) S_{p} \right] 	
		\right\rangle }
	     {\langle Q_{K} \rangle } +      \nonumber  \\ 
         & & \hspace{5mm} \left. \mbox{} +  J_{k'k}(t)  
	\frac{ \left\langle 
	\Tp \left[ a_{k', \si}^{\dagger}(t_{+}) \ak(t'_{+}) 
	         \csd(t'_{-})  c_{d, \si}(t_{-}) S_{p} \right] \right\rangle }
	{\langle Q_{K} \rangle } 	\right\}, \nonumber
\enea
Since the averages in this equation are taken in the full Hilbert space it is 
amenable for perturbation expansion. 

As in section \ref{Anderson} and Ref. \onlinecite{SivanWingreen96}, 
we assume that calculation of the tunneling current to 
lowest order in the tunneling strength 
which encodes the Kondo effect (i.e. $ J_{k'k}^{3} $) 
can be done using unperturbed populations of the 
energy levels. This assumption seems more natural here because the 
number of fermions in the dot is completely fixed by the constraint to 
the subspace $ F_{1}^{K} $. It is not the case in the Anderson model 
where slave fermions can convert into slave bosons and vise versa 
within the same subspace $ F_{1}^{A} $. Moreover, two levels with 
different spins have equal occupation numbers after 
performing an ensemble 
(thermal) averaging. Therefore, in the Kondo model this assumption 
concerns only occupation of different energy levels in the reservoirs. 
The denominator in equation (\ref{CurrGCE_K}) 
can therefore be easily calculated, that is,
\beq 						\label{denom_K}
  \langle Q_{K} \rangle  =  2 n_{d}  
		\stackrel{\lam \rightarrow \infty}{ \longrightarrow } 
		2 e^{ - \beta \lam}, 
\eneq
where $ n_{d} =  1 / \left[ \exp ( \beta \lam ) +1 \right]  $ 
is the Fermi function for the dot electrons 
in the grand - canonical ensemble. In fact, as 
far as the dot electrons are considered in the grand - canonical 
ensemble, they are not real particles any more. For example, one might 
notice that in the physical limit $ \lam \rightarrow \infty $ their Fermi 
function $ n_{d} $ tends to zero. They might be better called ``auxiliary 
fermions'' or ``slave fermions''. To avoid confusion with slave 
fermions of section \ref{Anderson} we prefer to refer to them hereafter 
as ``dot fermions''. 

\subsection{Diagrammatic expansion of the current}
In the following we formulate a diagrammatic technique to expand the 
numerator of Eq. (\ref{CurrGCE_K}) and obtain the current. We skip 
the detailed derivation and present only its 
main steps and then present an 
explicit formulation of the pertinent
diagrammatic rules. The strategy 
is to start from perturbation expansion 
of Eq. (\ref{CurrGCE_K}) on the closed time - path. Next, 
transformation of the resulting expression is performed,
first to a single - time branch and, 
then, to the
physical representations of the non-equilibrium Green 
functions. Time - translation invariance of the unperturbed 
Green functions allows us to use the Fourier transform (for every 
Green function separately) and write integrals 
in the energy representation. The 
rest of this procedure will be explained along the course of 
formulation below.

Equation (\ref{CurrGCE_K}) expresses the current in terms of a pair of 
two - particle Green functions. Therefore, in this section we prefer to 
draw the diagrams in the standard form excepted for Green functions 
in textbooks (see Fig. \ref{Newfig6}). Alternatively, we could 
connect all the external lines in one ``external vertex'' as we did in 
section \ref{Anderson}, thus, obtaining diagrams of a circular shape. 
The difference is, of course, purely superficial. It influences only the 
rules of drawing, i.e. the first item of the 
diagrammatic rules formulated 
below. 

Here are the rules for drawing the diagrams and writing down the 
corresponding analytical expressions: 
\begin{enumerate}
	\item  Draw a line corresponding to a propagator of a dot 
fermion. In Fig. \ref{Newfig6} we used a dashed line for its 
notation. The diagram must include only one such line, since a 
presence of two or more lines of this type implies $ Q_{K} \geq 2 $. It 
would produce contribution to the numerator of equation 
(\ref{CurrGCE_K}) of second or higher order in powers of 
$ \exp ( - \beta \lam ) $. The contribution to the current 
would then be of 
first or higher order in powers of $ \exp ( - \beta \lam ) $, which 
will vanish at $ \lam \rightarrow \infty $. 
Attach $ m - 1 $ points to the dot 
- fermion line, where $ m $ is the power of $ J_{k' k} $ which the 
contribution of the diagram being considered is supposed to have in the 
expression for the current. These points represent vertices. Connect 
them by lead - electron lines leaving two loose ends corresponding to 
the external operators 
$ a_{k'}^{\dagger} $ and $  a_{k} $ in Eq. (\ref{CurrGCE_K}). In 
Fig. \ref{Newfig6} we used solid lines to denote propagators of 
lead electrons. In the present work we have considered diagrams for 
$ m = 2 $ (Fig. \ref{Newfig6}C) and $ m = 3 $ (Fig. 
\ref{Newfig6}A,B,D). 
	\item Introduce the following $ 2 \times 2 $ (Keldysh) matrices 
for every lead -  
electron  and dot - fermion line respectively:
\beq						\label{matrices_K}
  g_{k \si}\epbr  \equiv  \left( \begin{array}{cc}            
			  0   &   g^{a}_{k \si}\epbr  \\                   
                      g^{r}_{k \si}\epbr &   g^{c}_{k \si}\epbr           
                         \end{array}      \right),  \ 
  \xi_{\si}\ombr  \equiv  \left( \begin{array}{cc}            
			  0   &   \xi^{a}_{\si}\ombr   \\             
                      \xi^{r}_{\si}\ombr &   \xi^{c}_{\si}\ombr           
                         \end{array}      \right). 
\eneq 
The indices ``r'', ``a'' and ``c'' denote the retarded, advanced 
and correlation Green functions which are explicitly defined as follows: 
\bea					\label{unperturbedGF_K} 
  g^{r(a)}_{k \si}\epbr & = &	
      \frac{1}{\epsilon - \ek \pm i \gam}  \\
  g^{c}_{k \si}\epbr & = &			   
      \left[ 1-2 f(\ek) \right]  \left[ g^{r}_{k \si}\epbr  - g^{a}_{k 
\si}\epbr \right]   \nonumber \\ 
  \xi^{r(a)}_{\si}\ombr  & = &			
      \frac{1}{\omega  - \lam \pm i \gam}  \nonumber \\ 
  \xi^{c}_{\si}\ombr  & = &			      
      \left[ 1 -  2 \exp ( -\beta \lam )  \right]  
	\left[ \xi^{r}_{\si}\ombr  - \xi^{a}_{\si}\ombr  \right] 
\nonumber 
\enea 
where $ f(\ek) = 1 / \left[ \exp ( \beta \ek ) +1 \right] $ is Fermi function 
for lead electrons, the factor $ \exp ( -\beta \lam ) $ is the limiting case 
of the dot - fermion Fermi function 
$ n_{d} =  1 / \left[ \exp ( \beta \lam ) +1 \right]  $ at large $ \lam $. 
As in section \ref{Anderson} the matrices (\ref{matrices_K}) emerge
as a result of the passage
from the closed time - path to the normal time path before 
the transformation to the energy representation is carried out. 
Note that the Lagrange multiplier 
$ \lam $ appears explicitly
in the definition of $ \xi^{r(a)} $ although consideration 
of energy conservation shows that, in fact, it could be absorbed into a 
shift of $ \om $ in all Green functions for the dot - fermions. 
	\item For every vertex, introduce the following tensor: 
\beq 					\label{vertices_K}
   \tilde{J}_{k_{out} k_{in}} 
   J_{s}\left( \frac{ W_{(k_{out} k_{in})} }{ \hbar \Omega } \right) 
   (-1)^{s}  \eta^{i' j'}_{i j}, 
\eneq 
where $ k_{in} $ and $ k_{out} $ are momenta of incoming and 
outgoing lead electrons respectively, 
and $ \tilde{J}_{k_{out} k_{in}} $ means a matrix element 
(given by equation (\ref{JWtilde})). Further, 
$ J_{s}\left( W_{(k_{out} k_{in})} / ( \hbar \Omega ) \right) $ are 
Bessel functions in which the order 
$ s $ is a number of photons absorbed in this vertex 
(clearly, it can be negative), and
\beq 						\label{Winout}
   W_{(k_{out} k_{in})}   =  \left\{ \begin{array}{ll}  
		W, 
	&	\mbox{if } k_{out} \in L, \; k_{in}\in R     \\ 
		-W, 
	&	\mbox{if } k_{out} \in R, \; k_{in}\in L     \\ 
		0,   
	&	\mbox{if } k_{out}, k_{in} \in L  
		\mbox{ or } k_{out}, k_{in} \in R      
			\end{array}   \right. . 
\eneq 
It is useful at this point to recall the origin of 
the appearance of the Bessel functions, starting from the
expansion of the time - dependent matrix elements 
$ J_{k_{out} k_{in}}(t) $ (see equations (\ref{JW}) and 
(\ref{Delta_K})): 
\begin{eqnarray}                            \label{Jexp_int} 
  \lefteqn{ J_{k_{out} k_{in}}(t)  =  \tilde{J}_{k_{out} k_{in}} 
	e^{ \frac{i}{\hbar} \left[  \phi_{(k_{out} k_{in})} t  + 
	i  W_{(k_{out} k_{in})}/ (\hbar \Omega ) 
	\sin \left( \Omega t + \alpha \right) 
	    -  \beta_{(k_{out} k_{in})} \right]  } } & &    \\
   & = & \tilde{J}_{k_{out} k_{in}} 
	\sum_{s=-\infty}^{\infty} 
            J_{s}\left( \frac{ W_{(k_{out} k_{in})} }
                                    { \hbar \Omega } \right) 
	(-1)^{s}  e^{ \frac{i}{\hbar} \left[ \phi_{(k_{out} k_{in})} t 
	- s \Omega t - s \alpha - \beta_{(k_{out} k_{in})} \right] } 
					\nonumber
\end{eqnarray}
The symbols $ (k_{out} k_{in}) $ here, as throughout the present 
manuscript, denote the leads to which
 $ k_{in} $ and $ k_{out} $ belong, that is, 
$ LR $, $ RL $, $ 
LL $ or $ RR $. The constant phases $ \beta_{(k_{out} k_{in})} $ 
cancel when equation (\ref{Jexp_int}) is substituted into expressions for 
the diagrams. 
The tensor $ \eta $ is now a tensor of the fourth rank (unlike in section 
\ref{Anderson}) because it connects four Keldysh matrices. It is written
explicitly as,
\beq					\label{eta_K}
   \eta^{i' j'}_{i j} = \sum_{m',n',m,n} 
			R_{i'm'}^{-1} R_{j'n'}^{-1} 
		\sigma_{m'n'}^{(3)} \delta_{m'm} \delta_{m'n} 
			R_{m i} R_{n j} , 
\eneq
where $ R = \frac{1}{\sqrt {2}}  
		       \left( \begin{array}{cc}            
				  1  &  1  \\                   
		                        -1  &  1          
                         		\end{array}      \right) $, 
$ R^{-1} = \frac{1}{\sqrt {2}}  
		       \left( \begin{array}{cc}            
				  1  & -1  \\                   
		                         1  &  1          
                         		\end{array}      \right) $, 
$ \sigma_{m'n'}^{(3)} =   
		       \left( \begin{array}{cc}            
				  1  &  0  \\                   
		                         0  &  -1          
                         		\end{array}      \right) $. 
The third Pauli matrix $ \sigma_{m'n'}^{(3)} $ appears here because 
of the transition from the closed - time path to the single - time 
representation (transformation from integration on the backward oriented
 time branch 
to integration over normal time axis requires a change of sign). 
Transformation to the physical representation employs the matrices $ R $ 
and $ R^{-1} $. 
	\item Close around the product of matrices and tensors which 
has been obtained by the following rows (or columns): 
$ 2^{-1/2} ( -1, 1)  $ for the $ g $ - matrices and 
$ 2^{-1/2} ( 1, 1)  $ for the $ \xi $ - matrices (see examples below). 
The origin of these rows is again the transformation from the single - 
time representation to the physical representation of the Green 
functions and vertices. Multiply the result by the factor, 
\beq
   \tilde{J}_{k' k} 
   J_{ n + s_{1}+ \ldots +s_{m-1} } 
	\left( \frac{ W_{(k' k)} }{ \hbar \Omega } \right) 
\eneq 
This factor emerges out of the matrix element $ J_{k'k}(t) $ in equation 
(\ref{CurrGCE_K}) by means of the expansion \ref{Jexp_int} (while $ 
s $ is replaced by $ -s $). Summation over $ n $ is carried out in equation 
(\ref{CurrDiaRes_K}) below. 
	\item Conserve energy in every vertex taking into account 
absorption of $ s $ photons and the energy $ \phi^{dc} $ gained (lost) 
by an electron going from left to right (from right to left). 
	\item Conservation of spin is somewhat delicate. There are two 
terms in the Hamiltonian (\ref{KondoFnl}) and, correspondingly, two 
terms in the expression for the current (\ref{CurrGCE_K}). They 
express spin - flip and normal scattering processes. Therefore, 
conservation of spin in every vertex should 
take into account two possibilities: 
\begin{enumerate} 
\item if incoming spins are opposite, then, spin flip occurs,  
\item if incoming spins are equal, then, both outgoing spins are the 
same.  
\end{enumerate} 
It is important to stress
that the same rules have to be applied to the whole 
diagram, otherwise it can 
not contribute to the current (\ref{CurrGCE_K}). Namely, 
\begin{enumerate} 
\item if incoming spins ($ \sigma_{e}' $ and $ \sigma_{f}' $ in Fig. 
\ref{Newfig6}) 
are opposite, then, the outgoing spins have to be reversed 
(i.e. $ \sigma_{e} = \sigma_{f}' = - \sigma_{e}' $ and 
$ \sigma_{f} = \sigma_{e}' = - \sigma_{f}' $), 
\item if incoming spins are equal, then, outgoing spins have to be the 
same  
(i.e. $ \sigma_{e} = \sigma_{f} = \sigma_{e}' = \sigma_{f}' $). 
\end{enumerate} 
These rules prohibit the spin - flip diagram of type B in Fig. 
\ref{Newfig6}. Indeed, spin - 
flip in the first vertex can happen only if 
$ \sigma_{e}' = - \sigma_{f}' $. Then, 
$ \sigma_{e,1} = - \sigma_{f,1} = - \sigma_{e}' = \sigma_{f}' $, and 
spin - flip should occur in the second vertex 
as well. Consequently, 
$ \sigma_{e} = - \sigma_{f} = \sigma_{e}' = - \sigma_{f}' $. 
This contribution is therefore
forbidden. The only diagram of type B which 
survives is the one 
with $ \sigma_{e} = \sigma_{f} = \sigma_{e,1} = \sigma_{f,1} =  
\sigma_{e}' = 
\sigma_{f}' $. On the other hand, four diagrams of type A are allowed. 
Namely, the following equalities $ \sigma_{e,1} = \sigma_{f,1} $, $ 
\sigma_{e}' = \sigma_{f} $ and 
$ \sigma_{f}' = \sigma_{e} $ should hold. All the combinations which 
are left are allowed, that is, 
$ \sigma_{e} = \pm \sigma_{f} $ and $ \sigma_{f,1} = \pm 
\sigma_{f} $. 
	\item Integrate over energy and momentum and sum over spin. 
Remember that, according to Eq. (\ref{CurrGCE_K}) the current flow 
starts from the left lead, i.e. $ k' \in L $, and ends in the right lead, i.e. 
$ k \in R $. The other momenta might belong to any lead: 
$ k_{1}, \ldots , k_{m-2} \in L,R $. Sum over numbers of absorbed 
photons 
$ s_{1}, \ldots , s_{m-1}$. 
	\item Multiply the result  by the factor 
\beq 
      \frac{i e}{ \hbar \langle Q_{K} \rangle } 
	(-i)^{m-1}  (i)^{2m}  (-1)^{F}  (2\pi )^{-(m+1)} 
\eneq  
where $ F $ is the number of closed electron - fermion loops. 
 The first factor in the above expression 
comes from equation (\ref{CurrGCE_K}), 
while the second one reminds us of the $ 
m-1 $ order of the expansion of the $S$ - matrix. 
Employing Wick's theorem in the perturbation series
requires pairings of creation - annihilation 
operators which are then expressed 
in terms of Green functions. This is the origin of the next
two factors. The last one emerges following
Fourier transforms. 
\item Take the 
limit 
$ \lam \rightarrow \infty $.
\end{enumerate} 
This completes the list of rules for 
drawing and calculating diagrams in the time - dependent Kondo model. 

\subsection{Explicit expression for the tunneling current}

Let us denote the expression obtained through the above rules for a 
certain diagram (or after summing of expressions for diagrams up to 
a certain order) as $ \Upsilon_{K}(n) $. The number $ n $ has the 
meaning of the total number of photons absorbed or emitted due to the 
tunneling process (it was introduced above in the forth item of the 
diagrammatic rules). The time - dependent current is then given by,
\bea					\label{CurrDiaRes_K}
  I(t) & = & \frac{1}{2} I_{0} +  
	\sum_{n=1}^{+\infty} \left| I_{n} \right| 
	  \cos (n \Omega t + n \alpha + \arg I_{n} ),   \\
  I_{n} & \equiv & \Upsilon_{K}(n) + \Upsilon_{K}^{*}(-n) .          
\nonumber   
\enea

As an example, an expression that is obtained using the above 
diagrammatic rules for diagram A in Fig. (\ref{Newfig6}) reads, 
\bea						\label{diaA_ruled}
\Upsilon_{K}^{A}(n) & = & - \frac{i e}{\hbar } 
	\sum_{\ssa, \ssb} 
	J_{n+\ssa+\ssb}\left( \frac{W}{\hbar \Omega} \right) 
	J_{\ssa}\left( \frac{W_{(\ka, k')}}{\hbar \Omega} \right)
	J_{\ssb}\left( \frac{W_{(k, \ka)}}{\hbar \Omega} \right)  
	(-1)^{\ssa+\ssb}                \nonumber  \\
& \cdot  &	\sum_{k' \in L; k \in R; \ka \in L,R; \si, \si', \sia} 
	\tilde{J}_{k'k} \tilde{J}_{k \ka} \tilde{J}_{\ka k'} 
	\frac{1}{4} \frac{1}{(2\pi )^{4}} 
	\int \!\! \int \!\! \int \!\! \int 
	            d \ea \, d\omega \, d\omega' \, d\oma \,  
	                                             \nonumber   \\ 
& \cdot  &  \lim_{\lam \rightarrow \infty } 
	\frac{1}{\langle Q_{K} \rangle}
	\sum_{i', j', i, j, \ia', \ldots, \jb} 
	(-1, 1)_{j} 
	g_{k \si'}^{j \jb'} 
	(\ea + \omega' - \oma + \phi_{(\ka k)} + \ssb \hbar  \Omega)
	                                          \nonumber   \\ 
& \cdot  &  \xi_{\sia}^{\ia \ib'}(\oma)   
	\eta^{\ib' \jb'}_{\ib \jb} 
	g_{\ka \sia}^{\jb \ja'}(\ea ) 
	\xi_{\si'}^{\ib i'}(\omega') 
	\left( \begin{array}{c}   1 \\  1  \end{array} \right)_{i'} 
	(1, 1)_{i} 
	\xi_{\si}^{i \ia'}(\omega)            \nonumber \\
& \cdot  &  \eta^{\ia' \ja'}_{\ia \ja} 
	g_{k' \si}^{\ja j'} 
	(\ea + \omega - \oma + \phi_{(\ka k')} - \ssa \hbar  \Omega)
	\left( \begin{array}{c}  -1 \\  1  \end{array} \right)_{j'} . 
\enea
A useful simplification can be worked out 
in this equation, noticing that one of 
the vertices connects electron Green functions of the same lead. 
Indeed, if $ \ka \in L $, then $ \ka $ and $ k' $ belong to the same lead, 
while if $ \ka \in R $, then $ \ka $ and $ k $ belong to the same lead. 
Therefore, according to Eq. (\ref{Winout}), one has  
$ W_{(k_{out} k_{in})} = 0 $ in this vertex. Using the identity, 
	$J_{s}(0) = \delta_{s,0}$,  
we can get rid of one of the Bessel functions. Using 
also the relation $J_{s}(-x)=(-1)^{s}J_{s}(x)$ we obtain,
\bea                                       \label{diaA_simpl}
\Upsilon_{K}^{A}(n) & = & - \frac{i e}{\hbar } 
	\sum_{s} 
	J_{n + s}\left( \frac{W}{\hbar \Omega} \right) 
	J_{s}\left( \frac{W}{\hbar \Omega} \right) 
	\sum_{k' \in L; k \in R; \ka \in L,R; \si, \si', \sia} 
	\tilde{J}_{k'k} \tilde{J}_{k \ka} \tilde{J}_{\ka k'} 
                                      \nonumber  \\
& \cdot  &  \lim_{\lam \rightarrow \infty } 
	\frac{1}{\langle Q_{K} \rangle}
	\frac{1}{4} \frac{1}{(2\pi )^{4}} 
	\int \!\! \int \!\! \int \!\! \int 
	d \ea \, d\omega \, d\omega' \, d\oma \,   \\ 
& \cdot  &	\sum_{i', j', i, j, \ia', \ldots, \jb} 
	 (-1, 1)_{j} 
	g_{k \si'}^{j \jb'} 
	(\ea + \omega' - \oma + \phi_{(\ka k)} + \ssb \hbar  \Omega)
	                          \nonumber   \\ 
& \cdot  &  \xi_{\sia}^{\ia \ib'}(\oma)   
	\eta^{\ib' \jb'}_{\ib \jb} 
	g_{\ka \sia}^{\jb \ja'}(\ea ) 
	\xi_{\si'}^{\ib i'}(\omega') 
	\left( \begin{array}{c}   1 \\  1  \end{array} \right)_{i'} 
	(1, 1)_{i} 
	\xi_{\si}^{i \ia'}(\omega)               \nonumber \\
& \cdot  &  \eta^{\ia' \ja'}_{\ia \ja} 
	g_{k' \si}^{\ja j'} 
	(\ea + \omega - \oma + \phi_{(\ka k')} - \ssa \hbar  \Omega)
	\left( \begin{array}{c}  -1 \\  1  \end{array} \right)_{j'} , 
\nonumber 
\enea
where $ s = \ssa $, if $ \ka \in R $, and $ s = \ssb $, if $ \ka \in L $. 

At this stage, the use of a program for analytical calculations (e.g. {\em 
Mathematica}) is indispensable. The following steps are to be done: 
(1) the tensor - matrix product is to be expanded and expressions for 
Green functions (\ref{unperturbedGF_K}) are to be substituted, 
(2) due to the presence of the factor $ \langle Q_{K} \rangle $ in the 
denominator, only terms of the first order in $ \exp (-\beta \lam ) $ must 
be kept in the numerator (the zero - th order cancel while the terms of 
higher orders are to be omitted), 
(3) routine integration of the expression obtained through the above 
steps over internal variables $ \ea $, $ \omega $, $ \omega' $ and $ 
\oma $ is to be carried out. The result is the following: 
\bea                                          \label{diaA_math}
\lefteqn{ \Upsilon_{K}^{A}(n)  =  \frac{i e}{2 \hbar }  
	\sum_{s} 
	J_{n + s}\left( \frac{W}{\hbar \Omega} \right) 
	J_{s}\left( \frac{W}{\hbar \Omega} \right)                
	\sum_{k' \in L; k \in R; \ka \in L,R; \si, \si', \sia} 
	\tilde{J}_{k'k} \tilde{J}_{k \ka} \tilde{J}_{\ka k'}  \cdot   }  \\
& &  \hspace{-1.5cm} \left[ 
	\frac{ f(\ek) f(\ekp) 
	   \left( \ek - \ekp - \phi^{dc} - s \Omega - 2 i \gam \right) }
	       { \left( \ek - \ekp - \phi^{dc} - s \Omega - i \gam \right) 
	   \left( \ek - \eka + \phi_{(k \ka)} - \ssb \Omega - i \gam \right) 
	   \left( \ekp - \eka + \phi_{(k' \ka)} - \ssa \Omega + 
                                  i \gam \right)  } - 
	                         \right.        \nonumber   \\ 
&  & \mbox{} - \frac{  f(\ek) f(\eka)  }
	   { \left( \ek - \ekp - \phi^{dc} - s \Omega - i \gam \right) 
	      \left( \ekp - \eka + \phi_{(k' \ka)} - \ssa \Omega + 
	                       i \gam \right)  } + 
	                                           \nonumber   \\ 
&  &  \left. \mbox{} + 
	\frac{ f(\ekp) f(\eka) }{
	 \left( \ek - \ekp - \phi^{dc} - s \Omega - i \gam \right) 
	 \left( \ek - \eka + \phi_{(k \ka)} - \ssb \Omega - i \gam \right)  } 
	\right]   .                       \nonumber  
\enea
We assume now that $ J_{k'k} $ depend only on the leads to which $ 
k' $ and $ k $ belong, independently of the values of $ k' $ and $ k $ 
(hence there are just three coupling - strength constants: 
$ J_{LL}$, $ J_{RR}$ and $ J_{LR}$, where 
$ J_{L(R)R(L)} = V_{L(R)} V^{*}_{R(L)} / | \ed | $). 
Furthermore, we assume that the densities of states in the leads 
($ \rho(\ek) $, $ k \in L,R $), are roughly constant on the energy scale of 
$ \phi^{dc} $, $ \Omega $ and $ W $ (rigorously speaking, only the 
combinations $ \rho(\ekp)^{1/2} \Jkpk \rho(\ek)^{1/2} $, where 
$ k, k' \in L,R $, need to be constant). 
Then the above expression greatly simplifies since integration over one 
of $ k $, $ k' $ or $ \ka $ can be carried out. Using also a simple 
relation, 
\beq
  \tilde{J}_{k'k} \tilde{J}_{k \ka} \tilde{J}_{\ka k'} = 
   \left| \tilde{J}_{LR} \right|^{2} \tilde{J}_{(\ka \ka)}, 
\eneq
we get an expression, 
\bea                                            \label{diaA_fnl}
\Upsilon_{K}^{A}(n) & = & \frac{e}{\hbar } 8 \pi 
	\left| \tilde{J}_{LR} \right|^{2} \rho_{L} \rho_{R} 
	\left( \tilde{J}_{LL} \rho_{L} + 
	        \tilde{J}_{RR} \rho_{R} \right)         \\ 
& \cdot &  \sum_{s} 
	J_{n + s}\left( \frac{W}{\hbar \Omega} \right) 
	J_{s}\left( \frac{W}{\hbar \Omega} \right) 
	\int \!\! \int d \ek \, d \ekp \,  
	\frac{ f(\ek) f(\ekp) }
	       { \ek - \ekp - \phi^{dc} - s \Omega - i \gam } . 
	                                            \nonumber  
\enea
The factor 8 in this equation arises from summation over spins. 

For the diagram B in Fig. \ref{Newfig6} the above diagrammatic 
rules, using $J_{s}(-x)=(-1)^{s}J_{s}(x)$ 
and $J_{s}(0)= \delta_{s,0}$, yield the following 
expression: 
\bea						\label{diaB_simpl}
\Upsilon_{K}^{B}(n) & = & \frac{i e}{\hbar } 
	\sum_{s} 
	J_{n + s}\left( \frac{W}{\hbar \Omega} \right) 
	J_{s}\left( \frac{W}{\hbar \Omega} \right) 
	\sum_{k' \in L; k \in R; \ka \in L,R; \si} 
	\tilde{J}_{k'k} \tilde{J}_{k \ka} \tilde{J}_{\ka k'}  
	                                        \nonumber  \\ 
& \cdot  &  \lim_{\lam \rightarrow \infty } 
	\frac{1}{\langle Q_{K} \rangle}
	\frac{1}{4} \frac{1}{(2\pi )^{4}} 
	\int \!\! \int \!\! \int \!\! \int d \ea \, 
	         d\omega \, d\omega' \, d\oma \,  \\ 
& \cdot  &	\sum_{i', j', i, j, \ia', \ldots, \jb} 
	 (-1, 1)_{j} 
	g_{k \si}^{j \jb'} 
	(\ea - \omega + \oma + \phi_{(\ka k)} + \ssb \hbar  \Omega)
	                                        \nonumber   \\ 
& \cdot  &  (1, 1)_{i} 
	\xi_{\si}^{i \ib'}(\omega) 
	\eta^{\ib' \jb'}_{\ib \jb} 
	g_{\ka \si}^{\jb \ja'}(\ea ) 
	\xi_{\si}^{\ib \ia'}(\oma) 
	\eta^{\ia' \ja'}_{\ia \ja}             \nonumber \\
& \cdot  &   g_{k' \si}^{\ja j'} 
	(\ea - \omega' + \oma + \phi_{(\ka k')} - \ssa \hbar  \Omega)
	\left( \begin{array}{c}  -1 \\  1  \end{array} \right)_{j'} 
	\xi_{\si}^{\ia i'}(\omega') 
	\left( \begin{array}{c}   1 \\  1  \end{array} \right)_{i'} . 
	                                        \nonumber  
\enea
Repeating the steps which led from Eq. (\ref{diaA_simpl}) to Eq. 
(\ref{diaA_fnl}) we obtain, 
\bea                                              \label{diaB_fnl}
\Upsilon_{K}^{B}(n) & = & \frac{e}{\hbar } 2 \pi 
	\left| \tilde{J}_{LR} \right|^{2} \rho_{L} \rho_{R} 
	\left( \tilde{J}_{LL} \rho_{L} + 
	        \tilde{J}_{RR} \rho_{R} \right)          \\ 
& \cdot &   \sum_{s} 
	J_{n + s}\left( \frac{W}{\hbar \Omega} \right) 
	J_{s}\left( \frac{W}{\hbar \Omega} \right) 
	\int \!\! \int d \ek \, d \ekp \,  
	\frac{ f(\ek) f(\ekp) }
	       { \ek - \ekp - \phi^{dc} - s \Omega - i \gam } . 
	                                     \nonumber  
\enea
We omitted in this expression some terms 
which cancel out when substituted 
into expression for the current (\ref{CurrDiaRes_K}). Note that spin 
- conservation rules discussed above allow 
only one 
combination of spins for this diagram, 
namely, all spin projections are identical. It leads to a 
factor 2 in the above expression instead of 8 in Eq. (\ref{diaA_fnl}). 

For the diagram C in Fig. \ref{Newfig6} we obtain, using the 
above diagrammatic rules and $J_{s}(-x)=(-1)^{s}J_{s}(x)$ 
an expression of second 
order in $ \Jt \rho $,   
\bea                                       \label{diaC_simpl}
\Upsilon_{K}^{C}(n) & = & \frac{i e}{\hbar } 
	\sum_{s} 
	J_{n + s}\left( \frac{W}{\hbar \Omega} \right) 
	J_{s}\left( \frac{W}{\hbar \Omega} \right) 
	\sum_{k' \in L; k \in R; \si, \sia} 
	\left|  \tilde{J}_{k'k} \right|^{2}  \nonumber  \\ 
& \cdot  &  \lim_{\lam \rightarrow \infty } 
	\frac{1}{\langle Q_{K} \rangle}
	\frac{1}{4} \frac{1}{(2\pi )^{3}} 
	\int \!\! \int \!\! \int \!\! \int d \eps' \, 
	                d\omega \, d\omega' \,  \\ 
& \cdot  &	\sum_{i', j', i, j, \ia', \ldots, \ja} 
	(-1, 1)_{j} 
	g_{k \si}^{j \ja'} 
	(\eps' - \omega + \omega' + \phi^{dc} + s \hbar  \Omega)
	                                       \nonumber   \\ 
& \cdot  &  (1, 1)_{i} 
	\xi_{\sia}^{i \ia'}(\omega) 
	\eta^{\ia' \ja'}_{\ia \ja} 
	g_{k' \sia}^{\ja j'} (\eps' ) 
	\left( \begin{array}{c}  -1 \\  1  \end{array} \right)_{j'} 
	\xi_{\si}^{\ia i'}(\omega') 
	\left( \begin{array}{c}   1 \\  1  \end{array} \right)_{i'} . 
                                                    \nonumber  
\enea
Making the same steps as above we obtain a very simple equation: 
\bea                                     \label{diaC_fnl}
\Upsilon_{K}^{C}(n) & = & \frac{e}{\hbar } 2 \pi 
	\left| \tilde{J}_{LR} \right|^{2} \rho_{L} \rho_{R}   \\ 
& \cdot &   \sum_{s} 
	J_{n + s}\left( \frac{W}{\hbar \Omega} \right) 
	J_{s}\left( \frac{W}{\hbar \Omega} \right) 
	\cdot \left( \phi^{dc} + s \hbar \Omega + i \  \mbox{const} 
	                          \right) .      \nonumber  
\enea
The last term (i.e. $ i \cdot \mbox{const} $) cancels when substituted 
into expression for the current (\ref{CurrDiaRes_K}). 

In order to calculate contribution of diagram D it is better to stay in the 
closed time - path representation. Then one can easily notice that it can 
be expressed through the contribution of diagram C in the following 
manner: 
\beq                           \label{diaD}
    \Upsilon_{K}^{D}(n) = 
	\Upsilon_{K}^{C}(n) \int_{p} d t_{p} 
	\sum_{\ka \in L,R} J_{\ka \ka}(t) n_{\ka} ,  
\eneq
where $ n_{\ka} \equiv \left\langle a_{\ka}^{\dagger} 
	a_{\ka} \right\rangle = f(\ek) $. 
The matrix elements $ J_{\ka \ka}(t) $ do not depend on the branch to 
which $ t $ belongs, i.e. whether $ t_{p} = t_{-} $ or $ t_{p} = t_{-} 
$, while $ n_{\ka} $ does not depend on time at all. So the integrals 
over the normal and the backward oriented time 
branches cancel. The contribution of 
this diagram to the current simply vanishes. 

Further progress employs the identities 
\bea                              \label{BesselSums}
    \sum_{s} J_{s}(x) J_{s+n}(x) & = & \delta_{n,0}      \\
    \sum_{s} s J_{s}(x) J_{s+n}(x) & = & 
	\frac{x}{2} ( \delta_{n,1} + \delta_{n,-1} ).   \nonumber 
\enea
Substituting Eq. (\ref{diaC_fnl}) into Eq. (\ref{CurrDiaRes_K}) and 
using these equations we obtain for the contribution of diagram C to the 
current ($ I^{(2)}(t) $) the following simple expression:  
\beq						\label{I2}
  I^{(2)}(t)  =  C_{2} \left[ \phi^{dc} + 
	W \cos (\Omega t + \alpha) \right] , 
\eneq
where $ C_{2} \equiv \frac{e}{\hbar} \pi 
	\left| \tilde{J}_{LR} \right|^{2} \rho_{L} \rho_{R} $. 
This is the only contribution which is of second order in $ \Jkpk 
\rho $. There appear only {\em DC} and the first harmonic. The higher 
harmonics are not generated in this order. 

Finally we sum contributions of all the diagrams and get the following 
equations for the tunneling current: 
\begin{mathletters}
\label{CurrFnlFormal}
\bea                
  I(t) & = & I^{(2)}(t) + I^{(3)}(t)       \\
  I^{(2)}(t) & = & C_{2} \left[ \phi^{dc} + 
	W \cos (\Omega t + \alpha) \right] ,       \\
  I^{(3)}(t) & = & \frac{1}{2} I_{0} +  
	\sum_{n=1}^{\infty} \left| I_{n} \right| 
	  \cos (n \Omega t + n \alpha + \arg I_{n} ),     \\
I_{n} & \equiv C_{3} & \sum_{s = -\infty}^{+\infty} \label{I_n_K}
	J_{s}\left( \frac{W}{\hbar \Omega} \right) 
	\left[ 
	 J_{s + n}\left( \frac{W}{\hbar \Omega} \right) 
	 F( \phi^{dc} + s \Omega, T, D )  + \right.         \\
	& & \hspace{3cm}  \left. \mbox{} + 
	 J_{s - n}\left( \frac{W}{\hbar \Omega} \right) 
	 F^{*}( \phi^{dc} + s \Omega, T, D )  \right],  
                                                                         \\
  F( \phi, T, D ) & = &                     \label{F_K}
	\int_{-D}^{+\infty} \!\! \int_{-D}^{+\infty} d \ek \, d \ekp \,  
	\frac{ f(\ek) f(\ekp) }{ \ek - \ekp - \phi - i \gam } ,   
\enea
\end{mathletters}
where $ C_{2} = \frac{e}{\hbar} \pi 
	\left| \tilde{J}_{LR} \right|^{2} \rho_{L} \rho_{R} $ and 
$ C_{3} \equiv \frac{e}{\hbar } 10 \pi 
	\left| \tilde{J}_{LR} \right|^{2} \rho_{L} \rho_{R} 
	\left( \tilde{J}_{LL} \rho_{L} + \tilde{J}_{RR} \rho_{R} 
\right) $, 
while $ \rho_{L(R)} $ are densities of states in the leads. The quantities 
$ I^{(2)} $ and $ I^{(3)} $ express contributions of second 
(diagram C) and third (diagrams A and B) orders in $ J_{k'k} \rho 
$ respectively. The cutoff $ D $ is equal to the energy difference 
between the chemical potential and the bottom of the conduction band, 
while $ \gam $ is an infinitesimally small number. 
Equations (\ref{CurrFnlFormal}) constitute the central formal result of the 
present section. 
Equation (\ref{F_K}) is rather convenient for further elaboration (which 
is our next step) but it can be 
misleading if a proper care is not taken. Indeed, the
cutoffs for the two energy integrations are not independent, 
and their relation is 
to be defined carefully. Consider another expression for $ F $ where 
one of the integrals does not require a cutoff: 
\beq					\label{F_LettForm}
  F( \phi, T, D )  =  - \frac{1}{2} \text{Re} 
		\int \!\! \int_{-D}^{+\infty} d\omega \, d\epsilon \,
		  \frac{ \left[ f_{L}(\omega) - f_{R}(\omega) \right] 
			\left[ f_{L}(\epsilon) + f_{R}(\epsilon) \right] }
		         {\omega - \epsilon + i \gamma } 
	-  i \pi \frac{\phi}{2} \coth \frac{\beta \phi}{2} , 
\eneq
Here $ f_{L}(\epsilon) \equiv 
1 / \left( \exp [ ( \epsilon - \phi )/ k T] + 1 \right) $ and 
$ f_{R}(\epsilon) \equiv 1 / \left( \exp [ \epsilon / k T] +1 \right) $ 
have the meaning of Fermi functions in the leads (the left lead being 
shifted by $ \phi $). Some pure imaginary terms, linear in $ \phi $, that 
do not contribute to the current have been
omitted in this expression. The 
above integral can be written in the same 
form as in the {\em DC} result of 
Sivan and Wingreen \cite{SivanWingreen96}. The
physical content of equation (\ref{I_n_K}) for $ I_{n}$ is
rather transparent: An alternating field applied to a two - barrier system 
actually results in a splitting of the leads energy levels 
\cite{TienGordon63}. Therefore the time - dependent current is a result 
of interference between ``{\em DC} - like'' contributions, 
each one of them having an 
effective bias $ \phi^{dc} + s \Omega $ weighted by the appropriate 
product of Bessel functions \cite{comOneLeadAlt}. 
An approximate evaluation of the double integral in the above equations 
is possible both for the linear ($ \phi \ll T $) and for the nonlinear 
($ \phi \gg T $) regimes. It yields,  
\beq				
				\label{Fapprox}
  \text{Re} [F( \phi, T, D )]   =  \left\{ \begin{array}{c}            
       \phi \left[ \ln \frac{D}{kT} + 0.26 + 
           O \left( \phi/kT \right) + O \left( kT/D \right) \right], 
\text{ if $ \phi \ll kT $},  \\                   
       \phi \left[ \ln \frac{D}{|\phi|} + 1 + 
           O \left( (kT/\phi)^{2} \right) + O \left( \phi/D \right) \right], 
\text{ if $ \phi \gg kT $}.   
                         \end{array}      \right.  
\eneq
As is well known (see, for instance, Ref. \onlinecite{Hewson:book93}), 
the linear - response perturbation expansion for the Kondo problem 
breaks down at temperatures below a certain value,
defined as the Kondo temperature $ T_{K} $. In our calculation it 
shows as a logarithmic divergence of the function $ F $ and, 
consequently, the amplitude of the current (\ref{I_n_K}) at $ T 
\rightarrow 0 $ in the linear response. It is remarkable, however, that in 
the nonlinear case on which we focus our attention here the function $ F 
$ does not diverge with decreasing temperature. Hence, our results for 
the nonlinear response are valid even below the Kondo temperature. 
This is due to the fact that here the non - linear bias plays the role of 
temperature as the largest low - energy scale. 

When the estimate (\ref{Fapprox}) is employed in equations 
(\ref{CurrFnlFormal}) it yields extremely simple expressions for the 
current. First, it turns out to be very useful for the static non-equilibrium 
case ($W=0$), for which we find,
\beq                                      \label{IdcApprox}
  I  \approx  \phi^{dc} \cdot \left[ \begin{array}{c}            
       C_{2} + 0.26 C_{3}+ C_{3} \ln \frac{D}{kT}, 
               \mbox{ if $ \phi^{dc} \ll T $}, \\                   
       C_{2} + C_{3} + C_{3} \ln \frac{D}{| \phi^{dc}|}, 
                    \mbox{ if $ \phi^{dc} \gg T $}.  
                         \end{array}      \right. 
\eneq
To the best of our knowledge, such a simple expression for a non - 
equilibrium tunneling current through a Kondo system has not been 
derived before. We notice that it contains the familiar pattern of the 
zero - bias anomaly (i.e. a peak in the differential conductance 
at zero bias). 

Next, considering expression (\ref{Fapprox}) we notice that the large 
factor $ \ln D $ appears only in a term which is linear in $ \phi $. Using 
equations (\ref{I_n_K}) and (\ref{BesselSums}), we then find that this 
factor emerges only in the expressions for the {\em DC} and the first 
harmonic of the {\em AC}. It means that, in fact, only the 
direct and the first harmonic are 
enhanced by the Kondo effect. As for the 
higher harmonics, the interference of the contributions to the current in 
equations (\ref{CurrFnlFormal}) with different effective biases is 
destructive. This conclusion is further confirmed by numerical 
calculation of the whole spectrum of the tunneling current performed 
using equations (\ref{CurrFnlFormal}). In Fig. \ref{Newfig7} the 
spectra (amplitude of the harmonics $ I_{n} $ via their number $ n $, 
the value of $ 1/2 I_{0} $ is displayed for 
the {\em DC}) are shown for both 
the Kondo and the non - interacting systems. Values of the parameters 
used for the calculation are listed in the caption. For the non - 
interacting system we used the equations of Ref. 
\onlinecite{GoldinAvishai97} reduced to the case of only one resonant 
level present in the system. Comparison of the top and bottom 
parts of the figure clearly shows a 
significant enhancement of {\em DC} and the first 
harmonic relative to the others in the Kondo system. 
The ratio of the {\em DC}
and the first harmonic to the higher harmonics might
increase even
further if $ D $ becomes larger. 
However, the choice of parameters for Fig. 
\ref{Newfig7} corresponds to a real quantum dot situation 
\cite{GoldhaberGordon98,Cronenwett98}. 
We emphasize that the non - linear response is different from the linear 
one. Namely, (i) the second and the higher harmonics exist although 
they are small, (ii) the amplitudes of the {\em DC} and the first harmonic are 
not determined any more by the ratio $ D/T $ as in the linear response, 
but rather by the ratio of $ D $ to the largest relevant low - energy scale 
as we show in the following (see equations (\ref{diffDCstrongAF}), 
(\ref{ACstrongAF}) and (\ref{ACweakAF}) below). 

Substituting equation (\ref{Fapprox}) into Eq. (\ref{CurrFnlFormal}), 
it is possible, in a few limiting cases, to find very simple expressions 
for the {\em DC} and the first harmonic of the time - dependent tunneling 
current. Experimentally, the nonlinear Kondo effect is usually revealed 
as the zero - bias anomaly. In a strong alternating field where many - photon 
absorption is possible (i.e. $ W \gg \Omega, T $), we get for the 
differential conductance 
($ \partial I^{dc} / \partial \phi^{dc} $) an expression, 
\bea                             \label{diffDCstrongAF} 
  \frac{ \partial I^{dc} }{ \partial \phi^{dc} } & \approx & 
      C_{2} + C_{3} \left[ \ln (D/W) + 1 \right]  + 
      C_{3} \frac{\Omega}{2W}  \sum_{|q|<W/\Omega} 
           P \left( \phi^{dc}/\Omega - q,  T/\Omega \right), 
           \text{ if } |  \phi^{dc} | \ll W,  \\
  \frac{ \partial I^{dc} }{ \partial \phi^{dc} } & \approx & 
      C_{2} + C_{3} \ln (D/| \phi^{dc} |),  
           \text{ if } |  \phi^{dc} | \gg W, \nonumber
\enea
where $ P \left( \phi^{dc}/\Omega - q , T/\Omega \right) $ is the shape 
function for the side peaks at $ \phi^{dc} = q \Omega $, $ q $ is 
integer. We find that, 
\bea                            \label{PfuncStrongAF}
  P \left( \phi^{dc}/\Omega - q,  T/\Omega \right) & \approx & 
      \frac{1}{2} 
      \ln \frac{ \Omega^{2} }{ (\phi^{dc} - q \Omega)^{2} +  T^{2} }, 
           \text{ if } |  \phi^{dc}/\Omega - q | \ll 1 
           \text{ and } |  T/\Omega | \ll 1 ,  \text{ while } \\
  P \left( \phi^{dc}/\Omega - q,  T/\Omega \right) & = & 0, 
           \text{ if } |  \phi^{dc}/\Omega - q | > 1 
           \text{ or } |  T/\Omega | > 1 . \nonumber
\enea
In a weak alternating field ($ W \ll \Omega $) , we obtain, 
 \beq                             \label{diffDCweakAF} 
  \frac{ \partial I^{dc} }{ \partial \phi^{dc} }  \approx  
      C_{2} + 
      \frac{C_{3}}{2} \ln \frac{D^{2}}{ (\phi^{dc})^{2} + T^{2} } + 
      C_{3}  \sum_{q \ne 0} 
               \left( \frac{W}{2\Omega} \right)^{ 2|q| }
               \frac{1}{ \left( |q|! \right)^{2} }
               P \left( \phi^{dc} - q \Omega, T, D \right), 
\eneq
where the side - peak shape function $ P $ is given by, 
\beq                            \label{PfuncWeakAF}
  P \left( \phi^{dc} - q \Omega, T, D \right)  \approx  
      \frac{1}{2} 
      \ln \frac{ D^{2} }{ (\phi^{dc} - q \Omega)^{2} +  T^{2} }. 
\eneq
Comparing equation (\ref{diffDCstrongAF}) with the static expression 
(\ref{IdcApprox}), we notice that the main peak of the differential 
conductance (the one at $ \phi^{dc} = 0 $) is suppressed in a strong 
alternating field by the factor approximately equal to 
$ \left[ \ln (D/W) + 1 \right] / \left[ \ln (D/T) + 0.26 \right]$. The side 
peaks in a strong field are not simple replicas of the 
central one, as is the case in a 
weak field. As far as the temperature dependence of the zero bias 
anomaly for the time - independent response is concerned, 
we note that the temperature 
dependence of the side peaks in a weak alternating field is governed 
by the factor $ \ln (D/T) $. In a strong alternating field, however, it is 
determined by the factor $ \ln (\Omega/T) $. 
Moreover, the amplitudes of the side peaks 
in a strong alternating field does 
not decay exponentially with $ q $. 
In fact, it is roughly constant (as long as $ 
q \Omega < W $). The half - width of the side peaks in a strong field 
is approximately 
equal to $ (\Omega T)^{1/2} $ instead of $ (D T)^{1/2} $. The latter 
feature is quite favorable 
for experimental observation: since the 
peaks are rather narrow it is not 
necessary to go to high frequencies in order to resolve 
them. On the other hand, their magnitude decreases 
together with the ratio $ \Omega/T $. In Fig. \ref{Newfig8} and 
Fig. \ref{Newfig9} the differential {\em DC} - conductance 
(calculated numerically using equations (\ref{CurrFnlFormal})) is 
shown versus the bias at different magnitudes of the alternating field. 
Suppression of the central peak in the zero bias anomaly
 is readily manifested in both 
figures. Side peaks do not appear in Fig. \ref{Newfig8} where the 
ratio of the frequency to the temperature is not large enough ($ 
\Omega/T = 5 $). They are well pronounced, however, at 
a tenfold lower 
temperature (Fig. \ref{Newfig9}). 

Although measurement of an {\em AC} with frequencies and 
amplitudes in the relevant range is not an easy task, it might reveal new 
interesting features of the Kondo effect. Kondo contribution to the 
direct tunneling current is usually revealed in an experiment through a 
special dependence on the parameters (such as $ \ln T $ increase of the 
conductance or the zero - bias anomaly). It was shown above that, as 
far as the spectrum of the tunneling current
is concerned, only the {\em DC} and the 
first harmonic are enhanced by the Kondo effect. This implies that such 
kind of parameter dependence can be found in the first harmonic as well 
as in the {\em DC} but not in higher harmonics. Employing the approximate 
expression (\ref{Fapprox}) in Eq. (\ref{CurrFnlFormal}) we obtain for 
the first harmonic of the time - dependent tunneling current ($ I^{ac} $) 
in a strong alternating field ($ W \gg \Omega $), 
\begin{mathletters}
\label{ACstrongAF}
\bea                                    
   I^{ac} & \approx &  C_{2}W + 
        \frac{2}{3} C_{3}W \left[ \ln (D/W) + 2 \right], 
           \text{ for } W \gg | \phi^{dc}|,T  \text{,}  \\
   I^{ac} & \approx &  C_{2}W + 
        \frac{2}{3} C_{3}W \left[ \ln (D/ | \phi^{dc}|) + 1/3 \right], 
           \text{ for } W,T \ll | \phi^{dc}|  \text{,}   \\
   I^{ac} & \approx &  C_{2}W + 
                       C_{3}W \left[ \ln (D/T) + 1/4 \right], 
           \text{for } W, | \phi^{dc}| \ll T.   
\enea
\end{mathletters}
These expressions appear to be very similar to equations 
(\ref{diffDCstrongAF}) for the {\em DC} when $ | \phi^{dc}| $ and 
$ W $ are exchanged. Moreover, differentiating $ I^{ac} $ with respect 
to $ W $, we find a peak in the {\em differential 
 AC - conductance} at 
zero {\em AC} - bias. The shape of the peak is logarithmic at large $ W $, 
while its height is determined by $ max(| \phi^{dc}|, T)$. 
We infer that this feature of the {\em AC} is analogous to the 
familiar zero bias anomaly
in the {\em DC}. In the same way that the latter is 
suppressed by the 
alternating - bias, this ``zero alternating - bias anomaly'' is 
suppressed by the 
direct - bias. In Fig. \ref{Newfig10} the differential 
alternating - conductance calculated 
using equations (\ref{CurrFnlFormal}) is 
plotted versus alternating
 - bias $ W $ at different values of the direct - bias $ 
\phi^{dc} $. Notice the clear similarity 
with the zero bias anomaly shown in Fig. 
\ref{Newfig8} and its suppression with 
alternating - bias. We do not 
find any side - peaks in the differential {\em AC} - conductance. 
Their traces
can be exposed, however, in the 
dependence of the {\em AC} on the 
direct - bias at low temperatures. 
In Fig. \ref{Newfig11} its derivative ($ 
\frac{dI^{ac}}{d\phi^{dc}} $) is drawn versus $ \phi^{dc} $ for two 
temperatures. At low temperature there appear well - 
pronounced dips (which are actually peaks in its absolute value) at 
integer multiples of the frequency. 

In a weak alternating field ($ W \ll \Omega $) we find, 
\begin{mathletters}
\label{ACweakAF}
\bea                              
   I^{ac} & \approx &  C_{2}W + C_{3}W \ln (D/ | \phi^{dc}| ),  
           \text{ for } T, \Omega \ll | \phi^{dc}|  \text{, }  \\
   I^{ac} & \approx &  C_{2}W + 
                                   C_{3}W \left[ \ln (D/ \Omega) + 1 \right],  
           \text{ for } T, | \phi^{dc}| \ll \Omega \text{ , } \\
   I^{ac} & \approx &  C_{2}W + 
                       C_{3}W \left[ \ln (D/T) + 1/4 \right],  
           \text{ if } | \phi^{dc}|, \Omega \ll T. 
\enea
\end{mathletters}
Inspecting equations (\ref{IdcApprox}), (\ref{diffDCstrongAF}), 
(\ref{ACstrongAF}) and (\ref{ACweakAF}) we notice that the values 
of both direct and alternating differential - conductances 
are basically determined 
by the logarithm of the ratio of $ D $ to the largest relevant low - energy 
scale. 

\section{Conclusions}

In the present work the problem of non-equilibrium time - dependent 
electron tunneling through an interacting system was 
studied at some depth. 
The main attention was
focused on calculation of time - dependent current in the 
Kondo regime beyond linear response. 
A tunneling system in this context is naturally 
described by the time - dependent Anderson model. 
Perturbation expansion of the current 
within this model, specially adapted for 
systems out of equilibrium was elaborated upon
in section \ref{Anderson}. This formalism combines the
non-equilibrium Green functions method with a specific 
approach suggested by Coleman \cite{Coleman84} 
to account for averaging in restricted subspaces 
which is often encountered in problems involving
strongly correlated electron systems. 
We have accomplished the formal part of perturbation expansion 
within this model for the time - dependent case. 
The task of performing detailed calculations 
turns out to be too formidable. Yet, with a slightly elevated 
capability of present day workstations and 
analytic software programs it should be feasible.

We then suggested a way to overcome 
this problem in section \ref{TDSW} where we developed
a time - dependent version of the Schrieffer - Wolff transformation 
mapping the time - dependent Anderson model onto a Kondo - type 
model. The latter is much easier for treatment within
perturbation theory. We maintain that it cannot 
be introduced phenomenologically since proper correlation of the time - 
dependence between the leads and the Kondo coupling constant 
$ J_{kk'}(t) $ has to be taken into account. Non-Equilibrium 
perturbation technique for calculation of the tunneling current within the 
time - dependent Kondo model was worked out in section 
\ref{KondoPerturbation}. Actual calculations were 
performed up to the third 
order in $ J_{kk'} $ (which corresponds to sixth order in tunneling 
matrix elements for the Anderson model) yielding extremely simple 
analytical expressions for the whole spectrum of the tunneling current 
(see equations (\ref{CurrFnlFormal}), (\ref{IdcApprox}), 
(\ref{diffDCstrongAF}), (\ref{ACstrongAF}) and 
(\ref{ACweakAF})). The nonlinear time - dependent current was found 
to be an interference sum of ``{\em DC} - like'' contributions, each one with an 
effective bias altered by the number of absorbed or emitted photons. We 
stress that our results are 
valid for the non - linear response both below 
and above the Kondo temperature although for the linear response they 
are valid only above it. 

There are three novel results in the present research. First, it was found 
that the Kondo effect strongly affects the first harmonic of the 
alternating tunneling current, 
no less that it affects the {\em DC}, while the 
other harmonics remain relatively small. This result 
is shown to originate from the 
interference of {\em DC} - like contributions to the current (equation 
(\ref{CurrFnlFormal}) that turn out to be rather destructive for all the 
harmonics except the {\em DC} and the first one. The higher harmonics are 
of course generated but 
their amplitudes are relatively small. This result was demonstrated to 
be remarkably different from that for a non - interacting one - level 
system where all the harmonics emerge together. 

Second, it was found that the zero - bias anomaly in the {\em DC} is
suppressed by an alternating field and 
displays side-peaks at multiples 
of the basic frequency. This result can be easily tested experimentally 
since it is concerned with measurement of the {\em DC} 
in the Kondo regime, which has now been 
well - established in quantum - dots experiments. Expressions 
(\ref{diffDCstrongAF}) together with Figs. \ref{Newfig8} and 
\ref{Newfig9} provide an estimate for the preferred range of 
parameters of the system. 

Third, we found a ``zero {\em AC} - bias anomaly'' in the alternating 
current, i.e. a peak of the differential {\em AC} - conductance at zero {\em AC} - bias. 
This phenomenon is an {\em AC} - analog of the familiar zero - bias anomaly 
of the {\em DC}. 
As in the latter one, it is suppressed by the {\em AC} - 
bias, while the former one is 
suppressed by the {\em DC} - bias. There are no side - 
peaks of the differential {\em AC} - conductance, however they have a well - 
pronounced counterpart in the derivative of the {\em AC} with respect to the {\em DC} 
- bias (in the form of dips). We think that this phenomenon shows a 
pattern of the Kondo effect in the {\em AC} yielding a 
challenging object for an experimental search. Equations 
(\ref{ACstrongAF}), (\ref{ACweakAF}) and Figs. 
\ref{Newfig10}, \ref{Newfig11} provide an estimate for the 
necessary range of parameters. Our results on the 
spectrum of the current indicate that 
effects like this one cannot appear at higher harmonics of the 
time - dependent tunneling current (due to the Kondo effect). 
We emphasize that expressions (\ref{diffDCstrongAF}), 
(\ref{ACstrongAF}) and (\ref{ACweakAF}), especially the numerical 
coefficients, are very approximate. They are intended to display the 
basic dependence of the current on the parameters of the system. 
Quantitative comparison with experiment may be done using the full 
set of equations (\ref{CurrFnlFormal}). 

 As far as relation to previous relevant works is concerned, we first 
notice that our analytical results for the {\em DC} are consistent with the 
numerical calculations of Ref. \onlinecite{HettlerSchoeller95}. 
However, being able to consider stronger {\em AC} - fields (larger ratio $ W / 
\Omega $), we find also an overall suppression of the zero - bias 
anomaly, beside the appearance of side peaks. As for the Fourier 
spectrum of the time - dependent tunneling current, we are unable to 
validate the assumption suggested in 
Ref. \onlinecite{HettlerSchoeller95} that 
{\em all} the harmonics beside the {\em DC} one can be neglected. Rather, the 
first harmonic is also enhanced, while the second and higher harmonics 
are generated but they are indeed much smaller than the {\em DC} and the first 
one \cite{comSpectrum}. Within a specific model, some authors 
\cite{SchillerHershfield96} obtained current spectrum similar to that of 
a non - interacting system. We attribute the difference between this 
result and ours to a quite peculiar choice of parameters used therein. 

We believe that further research on time - dependent aspects of the 
Kondo effect, in particular in quantum dots, is interesting and very 
timely. Let us mention a few possible directions of 
future research. First, the methods developed in
the present work can be adapted for solution of 
the problem of nonlinear response of a Kondo system to a combination 
of alternating magnetic and electric fields. Evolution of the zero - bias 
anomaly in a magnetic field, contrary to its temperature dependence, 
seems to be the clearest experimentally 
\cite{GoldhaberGordon98,RalphBuhrman94} resolved feature of the 
non-equilibrium Kondo effect. Calculation of 
complementary effects in the 
time - dependent response and 
carrying out pertinent experiments 
seem also to be timely. If the effect of a magnetic field can 
be solely expressed 
by the Zeeman splitting of the energy level in the dot, that is, 
$ \eps_{\si = \pm 1} = \eps_{0} \pm \Delta \eps /2 $, the technique 
developed in the present work can be easily modified to incorporate it. 

It might also be interesting to consider a non - magnetic (also called 
``orbital'') Kondo system \cite{Zawadowski80,VladarZawad83} 
subject to a strong alternating field. It is believed 
\cite{DelftRalph97,Cox98} to be realized in some recent experiments 
\cite{RalphBuhrman92,Keijsers9596} in the form of a two - level 
atomic tunneling system. In this model the conducting electrons 
interact with an impurity atom which can tunnel between two 
states. Tunneling of the atom is assisted by the interaction. At 
sufficiently low temperatures, the 
parameters of the model renormalize so 
that it becomes equivalent to the 2 - channel Kondo model 
\cite{Muramatsu86}. Besides the time - dependent shift of the leads 
which is present also in the 1 - channel model and studied in our work, 
in a two - level atomic system a time - dependent field can cause a 
change of the effective energy separation between the levels. One 
possible effect of such a change is a crossover between Fermi - liquid 
and non - Fermi - liquid behavior (see, for instance, Ref. 
\onlinecite{Cox98}). It is especially 
appropriate to point out here that application of an {\em AC} - 
field is rather controllable. An appropriate calculation, if followed by an 
experiment, could then further test the
hypothesis that the physics of a 2 - channel Kondo model
has been realized in some experiments,
\cite{RalphBuhrman92,Keijsers9596}, a point
which has been questioned
by some authors \cite{WingreenAltshMeir95}. 

Consideration of a multilevel Kondo system in an external alternating 
field looks very attractive. First, it is 
an appropriate object for quantum dots 
experiments. Moreover, it was argued \cite{Yamada84,Inoshita93} that 
the Kondo temperature in such a system can be enhanced by orders of 
magnitude. This fact could allow an experimental investigation of the 
strongly correlated regime of the Kondo system. On the other hand, 
application of a time - dependent field to a multilevel system leads to 
highly nonlinear tunneling processes, e.g. resonant frequency 
multiplication \cite{GoldinAvishai97}. Examination of non-equilibrium 
transport through such a system in the Kondo regime might reveal new 
and interesting effects. 

In closing, we believe that the physics of strongly correlated 
particles in strong time - dependent external fields and restricted 
geometries looks to be an important and exciting subject for further 
research. 

\noindent
{\bf Acknowledgments}:
This research is supported in part by a grant from the Israel 
Science Foundation
under programs {\em Centers of Excellence}, and {\em 
Non-Linear Tunneling}, by an American 
- Israel BSF grant {\em Dynamical Instabilities} 
and by a DIP program {\em Quantum Electronics in Low Dimensional 
Systems}. 
We would like to thank N.S. Wingreen, A. Golub, Y. Meir, D. 
Goldhaber - Gordon, L.P. Kouwenhoven, P. Coleman, L.I. Glazman, 
A. Schiller, K.A. Kikoin, G. Sch\"{o}n and E. Kogan for helpful 
discussions and comments.

\begin{figure}[htbp]
\caption{%
a) Schematic drawing of the Kondo peaks in the non-equilibrium 
interacting density of states of the dot (impurity). Here $ \epsilon_{d} $ 
is the bare energy level of the dot, while $ \mu_{L(R)} $ are chemical 
potentials in the left (right) lead. 
b) Possible formation of numerous Kondo - resonance peaks in the 
density of states caused by a time - dependent field. The peaks 
associated with tunneling to the left and the right leads are schematically 
shown near the left and the right barriers respectively. $ \Omega $ is the 
frequency of the external field. 
\label{Newfig1}}
\caption{Choice of points on the closed time - path for the external 
operators in equation (\protect\ref{Curr-CTPath_And}). 
\label{Newfig2}}
\caption{Diagrams for the perturbation expansion of the current in the 
Anderson model. Dashed lines stand for slave fermions, dash - dotted 
lines --- for slave bosons, solid lines --- for lead electrons. A quantity 
$ s_{i} $ is the number of photons emitted when an electron goes from 
the dot to a lead, while  $ q_{i} $ is the number of photons absorbed by 
an electron going from a lead to the dot. 
\label{Newfig3}}
\caption{Tunneling vertices in the slave - bosons representation of the 
Anderson model. 
(a) Tunneling of an electron from the dot to a lead is represented as the 
decay of a slave fermion into a slave boson and a lead electron. 
(b) Reverse process representing tunneling from a lead to the dot. 
\label{Newfig4}}
\caption{Choice of points on the closed time - path for the external 
operators in equation (\protect\ref{Curr-CTPath_Kondo}). 
\label{Newfig5}}
\caption{Diagrams for the perturbation expansion of the current in the 
Kondo model. Solid lines stand for lead electrons, dashed lines --- for 
dot electrons. 
\label{Newfig6}}
\caption{a) Spectrum of the tunneling current for a quantum dot in the 
Kondo regime (amplitude of the harmonics $ I_{n} $ via their number 
$ n $, the value of $ 1/2 I_{0} $ is shown for the dc).
The contribution of the second order in $ J \rho $ is not shown.
The current is measured in units of $ C_{3} \cdot k T $.
$ W /\Omega = 4 $, $ \Omega/ k T = 5 $, $ \phi^{dc}/ k T = 10 $, 
$ D/ k T = 200 $. 
b) Spectrum of the tunneling current for a non - interacting
one - level system with the energy level between the chemical
potentials of the left and the right leads.
The current is measured in units of $ e \Gamma / \hbar $, where
$ \Gamma $ is the level width.
$ W /\Omega = 4 $, $ \Omega/ \Gamma = 5 $, 
$ \phi^{dc}/ \Gamma = 10 $, $ D/ \Gamma = 200 $. 
\label{Newfig7}}
\caption{Differential {\em DC} - 
conductance (in units of $ C_{3} $) versus {\em DC} 
- bias at various values of {\em AC} - 
bias $ W $ (both are measured in units of 
$ k T $). $ \Omega/ k T = 5 $, $ D/ k T = 200 $. 
\label{Newfig8}}
\caption{The same as in the previous plot but at ten - fold lower 
temperature. 
\label{Newfig9}}
\caption{Differential {\em AC} - conductance (in units of $ C_{3} $) versus 
{\em AC} - bias at various values of {\em DC} - bias $ \phi^{dc} $ (both are 
measured in units of $ k T $). $ \Omega/ k T = 5 $, $ D/ k T = 200 $. 
\label{Newfig10}}
\caption{Derivative of the {\em AC} with respect to {\em DC} 
- bias versus the latter 
(which is given in units of $ \hbar \Omega $). $ W /\Omega = 2 $, 
$ D/ \Omega = 40 $. Dashed line is plotted at $ \Omega / T = 5 $, while 
solid line is at the ten - fold lower temperature. 
\label{Newfig11}}
\end{figure}


\begin{thebibliography}{99}
\bibitem{comOneLeadAlt} At moderate external fields (this condition is 
defined by inequalities (\ref{ExtFcond_small})), behavior of the Kondo 
system depends only on the potential difference between the leads, as 
discussed in the previous section. On the other hand, an application of a 
uniform (though time - dependent) potential is not observable even if it 
is arbitrary strong and arbitrary fast \cite{GoldinAvishai97}. Therefore, 
the time - dependent shift of the right lead can be subtracted and only 
the left lead is considered alternating.
\bibitem{comSpectrum} It is true also below the Kondo temperature. 
We explained it in the text for the nonlinear case. In the linear response 
the function $ F( \phi, T, D ) $ is linear in $ \phi $ by definition. Then it 
can be exactly proved that only the {\em DC} and the first harmonic are 
generated.
\end{thebibliography}
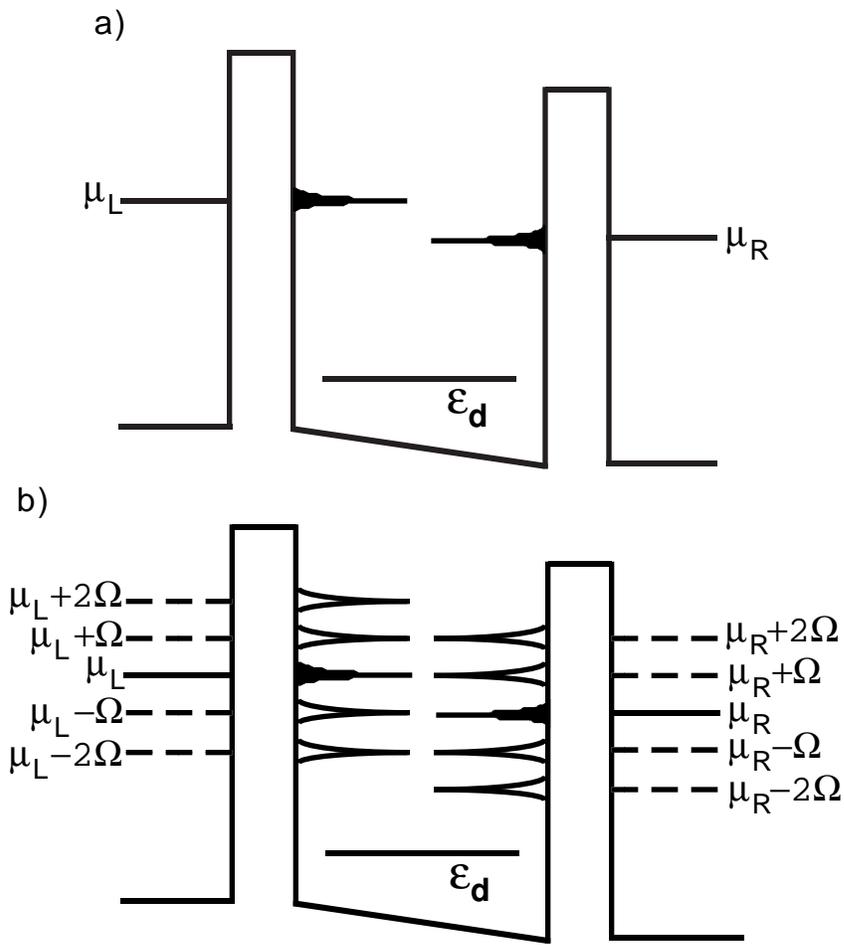
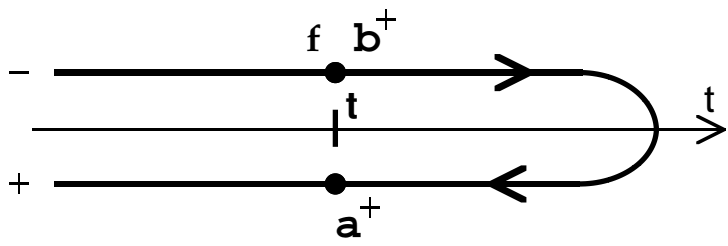
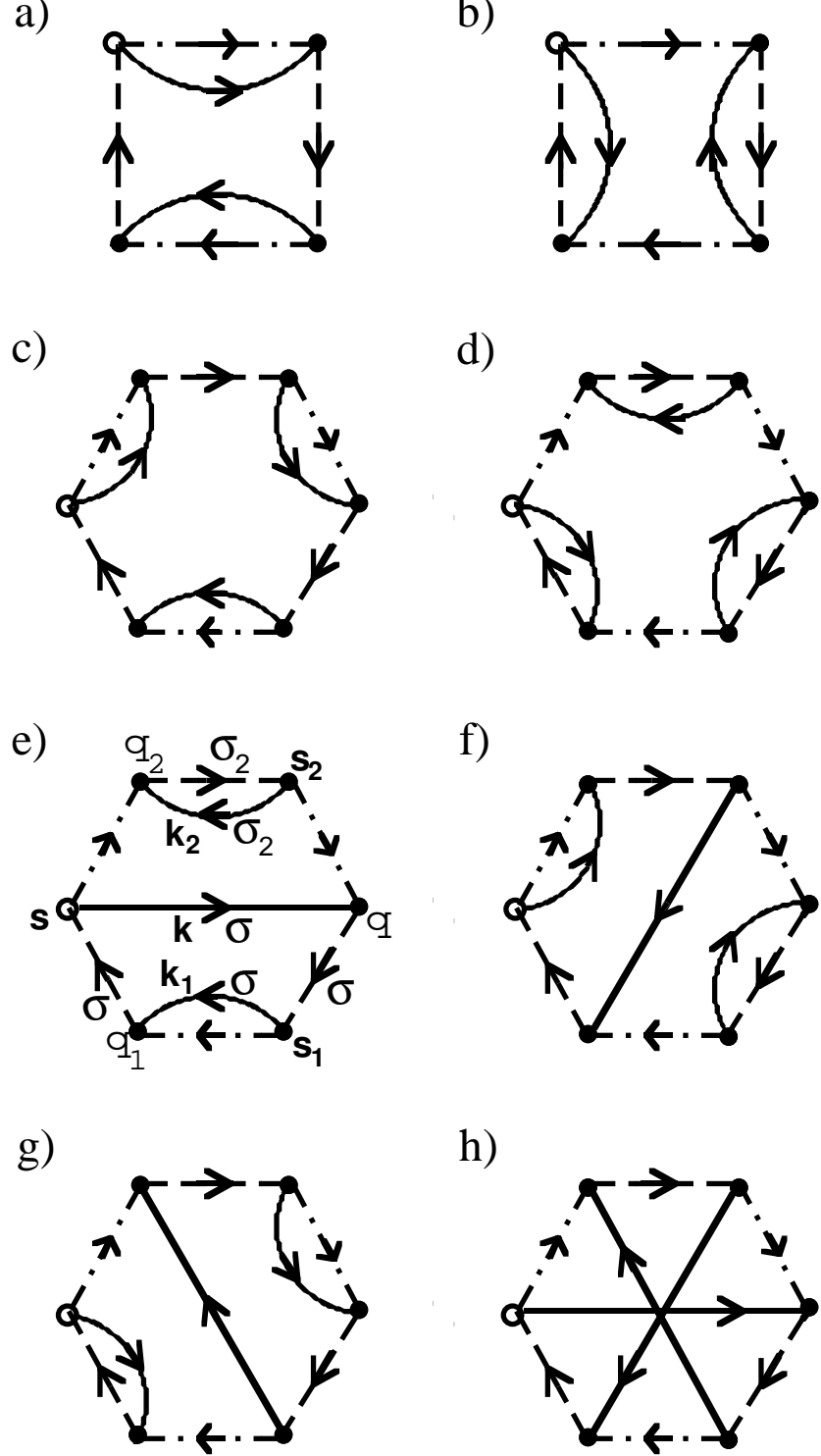
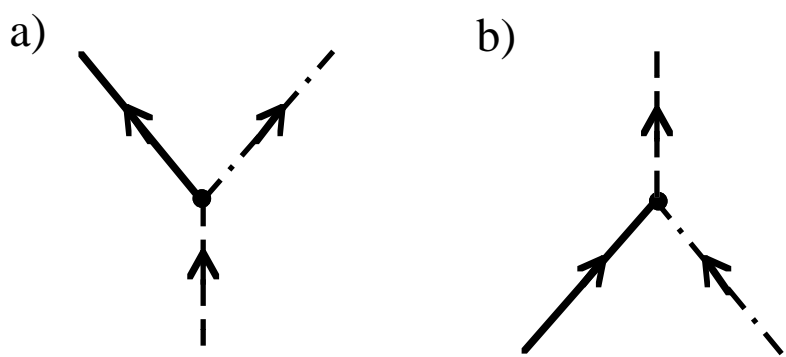
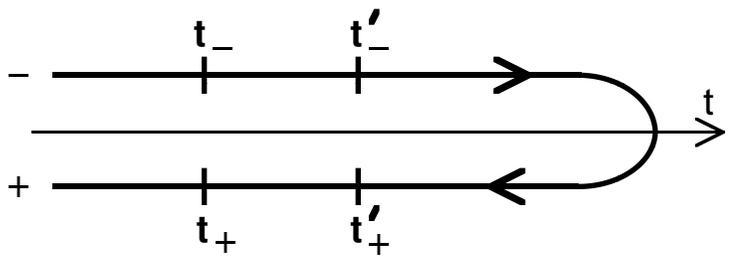
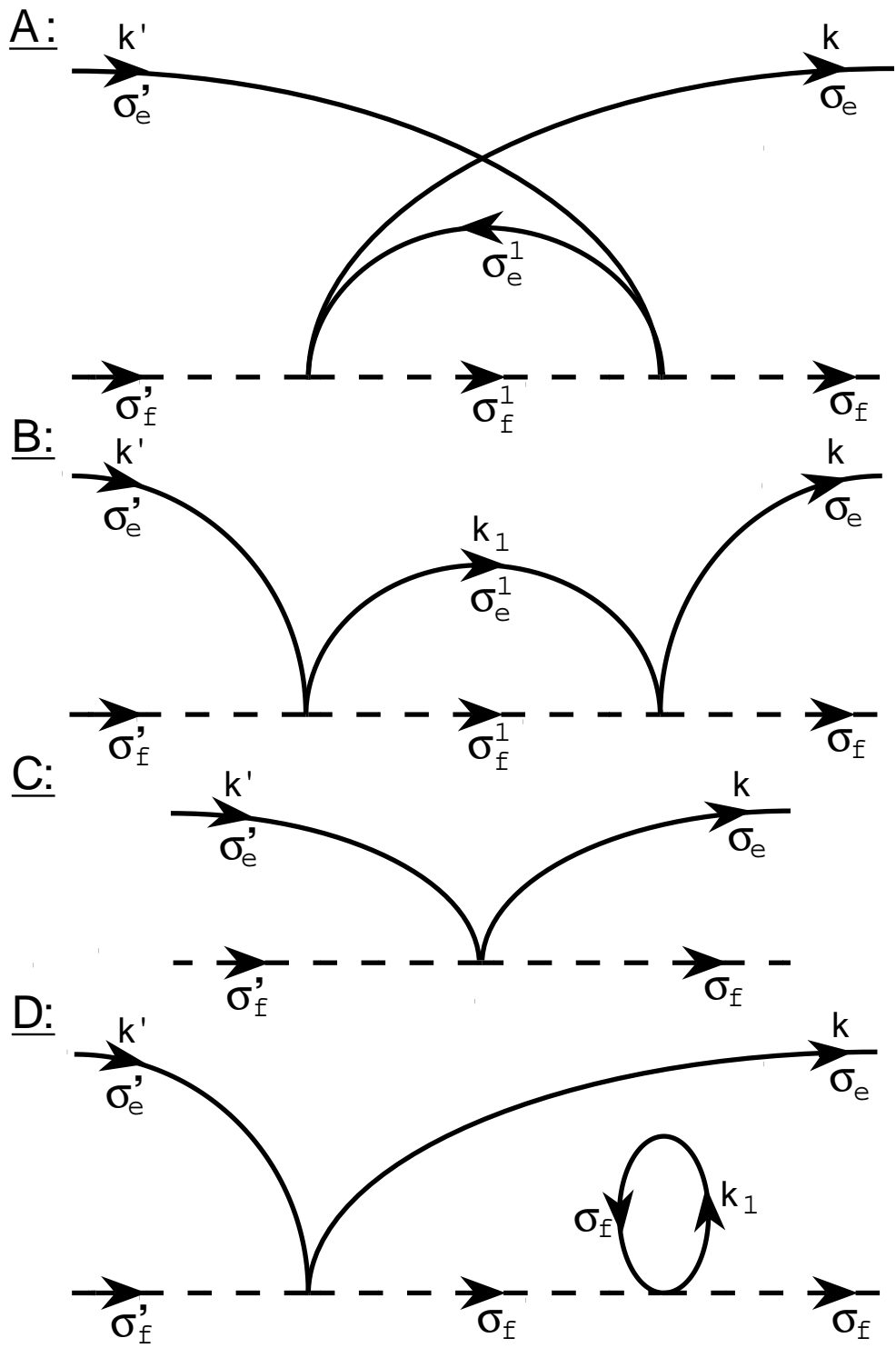
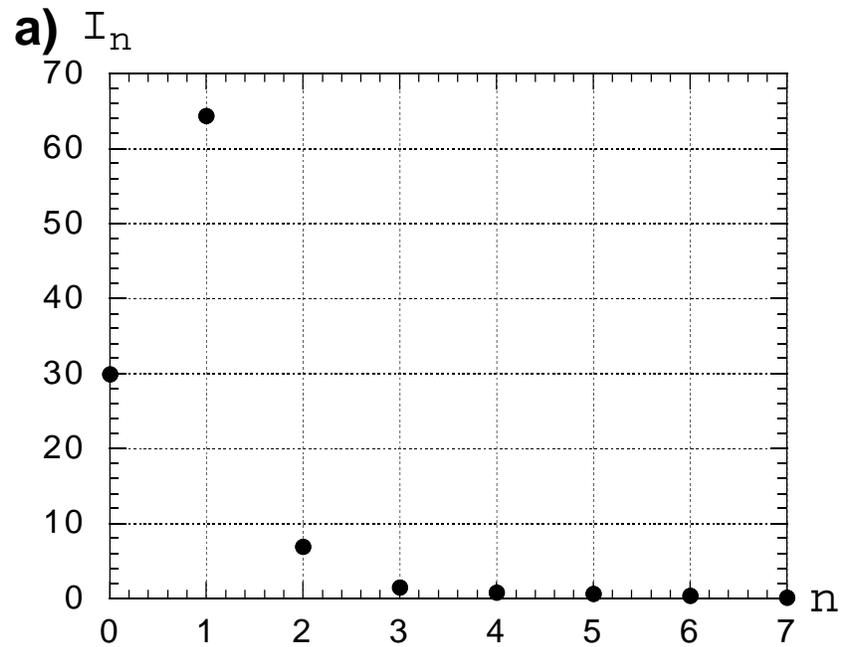
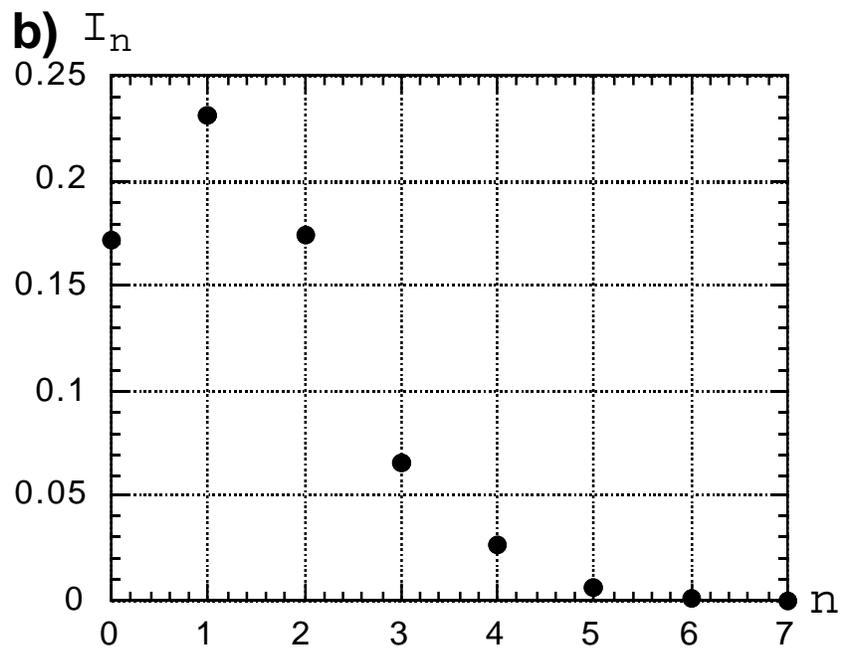
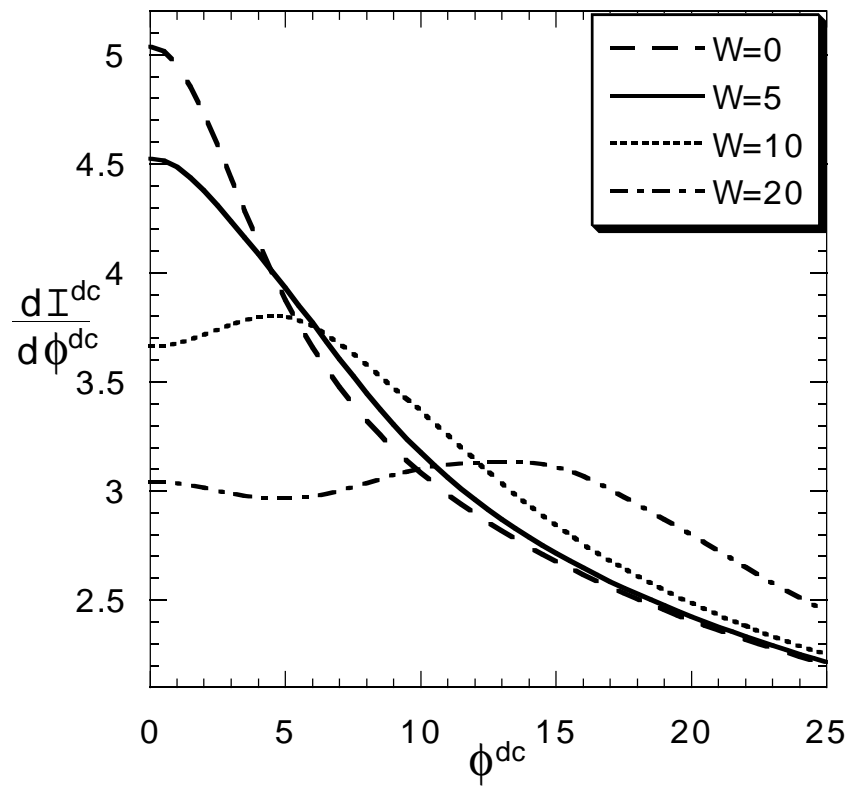
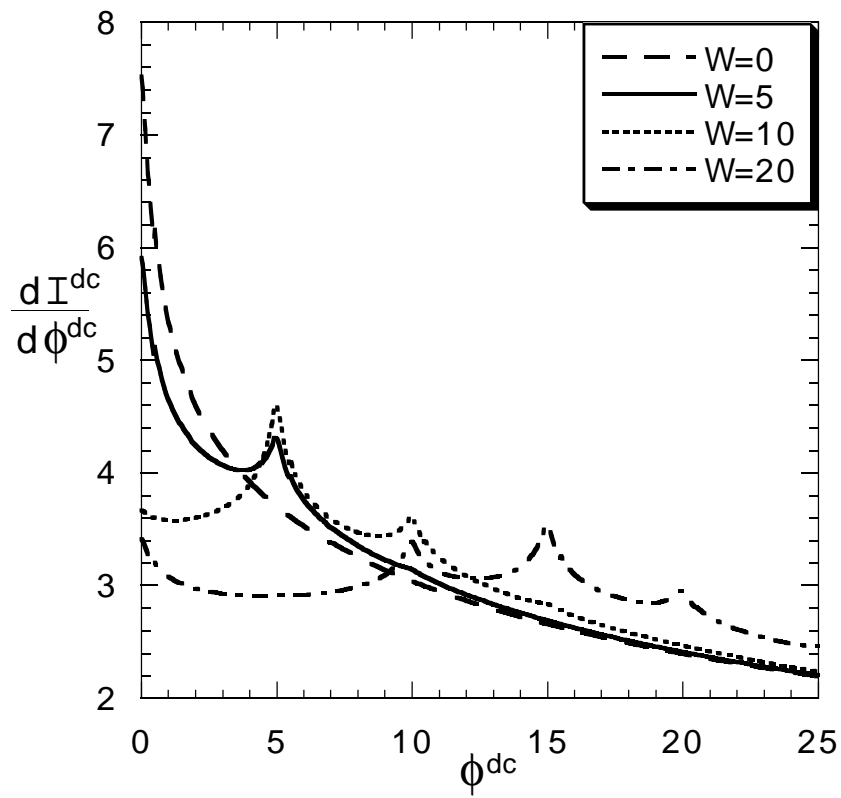
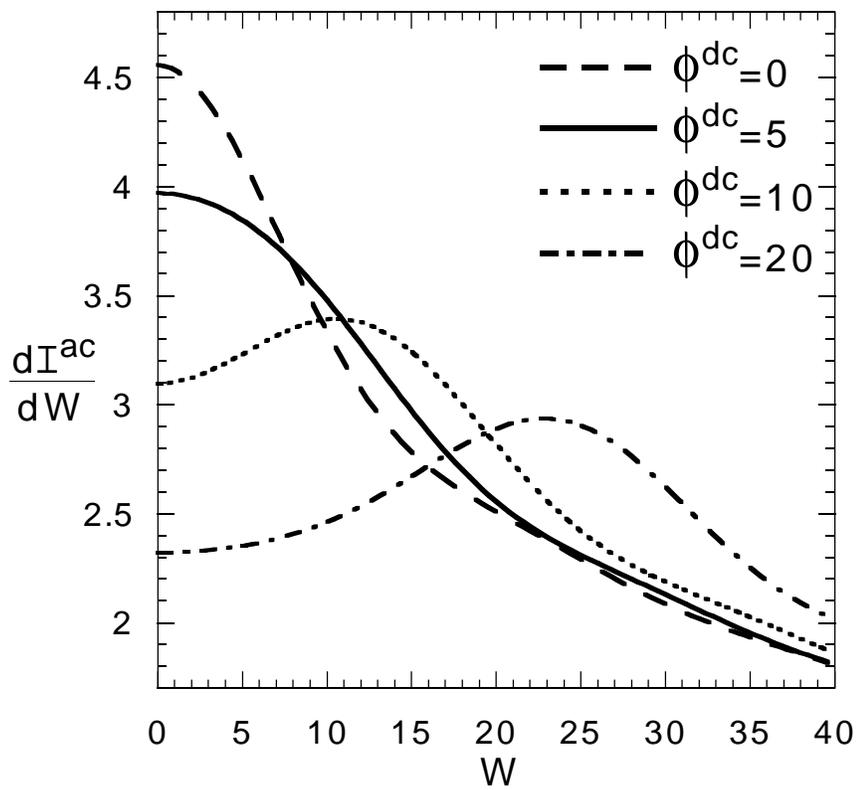
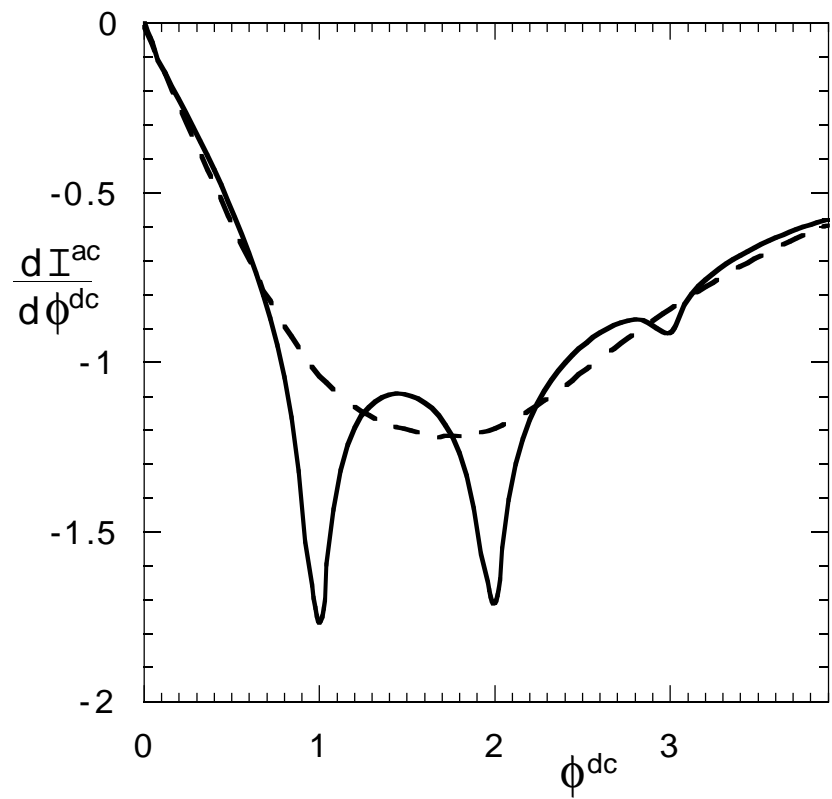
\end{document}